\documentclass[11pt]{article}
\usepackage{epsfig}
\usepackage{amsfonts}
\usepackage{amsmath}
\usepackage{amssymb}

\newcommand{\simle}{\hspace*{0.2em}\raisebox{0.5ex}{$<$}
     \hspace{-0.8em}\raisebox{-0.3em}{$\sim$}\hspace*{0.2em}}
\newcommand{\ep}{\epsilon}

\newcommand{\g}{\gamma}

\newcommand{\la}{\lambda}
\newcommand{\simu}{\sigma^{\mu\nu}}
\newcommand{\Fmu}{F_{\mu\nu}}
\newcommand{\Gmu}{G^a_{\mu\nu}}

\newcommand{\slashT}{\slash\hspace{-0.4em}T}

\newcommand{\slashP}{\slash\hspace{-0.6em}P}
\newcommand{\slashPsub}{\slash\hspace{-0.5em}P}
\newcommand{\slashPT}{\slash\hspace{-0.6em}P\slash\hspace{-0.4em}T}
\newcommand{\slashPTsub}{\slash\hspace{-0.45em}P\slash\hspace{-0.4em}T}

\newcommand{\qb}{\bar q}
\newcommand{\Nb}{\bar N}

\newcommand{\tb}{\bar \theta}

\newcommand{\MQCD}{M_{\mathrm{QCD}}}
\newcommand{\MNN}{M_{N\!N}}

\newcommand{\Or}{\mathcal O}

\newcommand{\dslash}[1]{#1 \llap{/\kern-0.5pt}}
\newcommand{\Dslash}[1]{#1 \llap{/\kern+1.2pt}}
\newcommand{\DDslash}[1]{#1 \llap{/\kern+2.3pt}}
\newcommand{\dslashh}[1]{#1 \llap{/\kern+1pt}}

\newcommand{\boldtau}{\mbox{\boldmath $\tau$}}
\newcommand{\boldpi}{\mbox{\boldmath $\pi$}}

\def\slashchar#1{\setbox0=\hbox{$#1$}           
  \dimen0=\wd0                                    
  \setbox1=\hbox{/} \dimen1=\wd1                  
  \ifdim\dimen0>\dimen1                           
    \rlap{\hbox to \dimen0{\hfil/\hfil}}            
    #1                                             
  \else                                          
    \rlap{\hbox to \dimen1{\hfil$#1$\hfil}}        
    /                                           
 \fi}                                           %

\begin{document}

\begin{titlepage}

\vspace{2.0cm}

\begin{center}

{\Large\bf 
Toroidal Quadrupole Form Factor of the Deuteron}

\vspace{1.5cm}

{\large {\bf E. Mereghetti$^{1}$, J. de Vries$^{2,3,4}$, 
R. G. E. Timmermans$^4$,} and {\bf U. van Kolck$^{5,6}$}} 

\vspace{0.5cm}

{\large 
$^1$ 
{\it Ernest Orlando Lawrence Berkeley National Laboratory,
University of California, \\ 
Berkeley, CA 94720, USA}}

\vspace{0.25cm}
{\large 
$^2$ 
{\it Institute for Advanced Simulation, Institut f\"ur Kernphysik, and \\
J\"ulich Center for Hadron Physics, Forschungszentrum J\"ulich, \\
D-52428 J\"ulich, Germany}}

\vspace{0.25cm}
{\large 
$^3$ 
{\it Nikhef, Science Park 105, \\ 
1098 XG Amsterdam, The Netherlands}}

\vspace{0.25cm}
{\large 
$^4$ 
{\it KVI, University of Groningen, \\ 
9747 AA Groningen, The Netherlands}}

\vspace{0.25cm}
{\large 
$^5$ 
{\it Institut de Physique Nucl\'{e}aire, 
Universit\'e Paris Sud, CNRS/IN2P3,\\
91406 Orsay, France}}

\vspace{0.25cm}
{\large 
$^6$ 
{\it Department of Physics, University of Arizona,\\
Tucson, AZ 85721, USA}}

\end{center}

\vspace{1.5cm}

\begin{abstract}
We calculate the toroidal quadrupole moment and form factor of the
deuteron, which violate time-reversal symmetry but conserve parity, at
leading order in two-flavor chiral effective field theory with perturbative
pion exchange. We take into account time-reversal and parity violation
due to the QCD vacuum angle combined with parity violation resulting
from the weak interaction in the Standard Model. We also consider time-reversal
and parity violation that at the quark-gluon level results from effective
dimension-six operators originating from physics beyond the Standard Model.
\end{abstract}

\vfill
\end{titlepage}

\section{Introduction}
It has long been known that particles with non-zero spin can have
``toroidal'' electromagnetic  form factors that are odd under charge
conjugation $C$, which implies that they violate either parity ($P$)
or time reversal ($T$), but not both symmetries simultaneously~\cite{Khr97}.
The toroidal dipole form factor (TDFF), also called the anapole 
\cite{dianapole},
requires spin 1/2 or higher, violates $P$ and conserves $T$.
The toroidal quadrupole form factor (TQFF), which requires spin 1 or higher,
violates $T$ and conserves $P$, and so on \cite{quadruanapole}. 
Toroidal form factors produce no physical effects when the photon is
on-shell, and correspond in a classical picture to fields 
within the charge distribution~\cite{Gra10}. 
These features contrast with the
more familiar $C$-even
electric and magnetic form factors, which respect or violate both 
$P$ and $T$ simultaneously, and produce effects for on-shell photons. 
The only form factors allowed for massive particles that are their
own antiparticles are toroidal \cite{boudjema}.

The toroidal form factors do contribute to the short-range 
interaction with a charged particle.
For nucleons and nuclei, in particular, 
they are in principle accessible via lepton scattering.
While there exist calculations of the TDFFs of the nucleon
and nuclei, there is apparently no calculation of a nuclear TQFF.
The TQFF of positronium was calculated in Ref. \cite{atomic}.

The aim of this paper is to provide the first controlled calculation of the
TQFF of the simplest nucleus, the deuteron, at low momentum.
The Lorentz-covariant electromagnetic current of a particle with spin $1$ is 
described by seven electromagnetic form factors:
charge, magnetic dipole, and electric quadrupole, which
are $P$- and $T$-conserving ($PT$);
electric dipole and magnetic quadrupole, which 
are $P$- and $T$-violating $(\slashPT)$;
TDFF, which is $P$-violating and $T$-conserving ($\slashP T$);
and, finally, 
TQFF, which is $P$-conserving and $T$-violating $(P\slashT)$.
We can write the spatial, $P \slashT$ component of the electromagnetic current 
as \cite{quadruanapole}
\begin{equation}
\langle \vec p^{\,\prime}, j | J^{k}_{P\slashT} | \vec p, i \rangle = 
i \left[ q^i q^j q^k 
- \frac{\vec q^{\, 2}}{2} \left(\delta^{i k} q^j + \delta^{j k} q^i \right)\right] 
F_{P \slashT}(\vec q^{\, 2}),
\label{PCTVFF}
\end{equation}
where $| \vec p, i \rangle$ is a deuteron state with
momentum $\vec p$ and polarization $\delta^\mu_i$ in the rest frame,
normalized so that 
$\langle \vec p^{\,\prime}, j | \vec p, i \rangle =
\sqrt{1+{\vec p}^{\,2}/m_d^2}\, (2\pi)^3 \delta^{(3)}(\vec q)\delta_{ij}$,
$\vec q= \vec p - \vec p^{\,\prime}$ is the (outgoing) momentum of the photon,
$m_d=2m_N -\gamma^2/m_N +\ldots$ is the deuteron mass
in terms of the nucleon mass 
$m_N\simeq 940$ MeV and the binding momentum 
$\gamma \simeq 45$ MeV.
$F_{P \slashT}(\vec q^{\, 2})$ is the TQFF,
which is proportional to the proton charge
$e=\sqrt{4\pi \alpha_{\mathrm{em}}}$ and has dimensions of mass$^{-3}$.
We express it in units of $e$ fm$^3$.

We denote the corresponding toroidal quadrupole moment (TQM) by
$\mathcal T_d = F_{P \slashT}(0)$.
It can be viewed as an interaction of the deuteron $d$ with the 
second derivative of the magnetic field $\vec{B}$ of the form
\begin{equation}
\mathcal L = 
\frac{\mathcal T_d }{2} \,
d^{\dagger} \left\{ S_{i}, S_{j} \right\}d \
\nabla_i \left(\vec{\nabla} \times \vec B\right)_j,
\end{equation}
where $S$ denotes the deuteron spin, and $\{ .\, , \, .\}$ the anticommutator.
Using Maxwell's equations to replace the curl of the magnetic field with 
a current, we can trade the $P \slashT$ moment for a contact interaction. 
For example, 
the $P\slashT$ interaction of a non-relativistic lepton of mass $m_l$ 
with the deuteron becomes a dimension-eight contact interaction,
\begin{equation}
V = \frac{e \mathcal T_d }{2 m_l} \left\{ S_{i},  S_{j} \right\}
\left[\left(\nabla_i \delta^{(3)}(\vec x)\right) \hat{p}_j  
+ 
\epsilon_{i k m} \sigma_{k} \, \nabla_{m} \nabla_{j}
\delta^{(3)}(\vec x)\right].
\end{equation}
The first term is due to the lepton kinetic term 
and gives rise to a non-local interaction involving $\hat{p}=-i{\vec\nabla}$. 
The second one comes from the interaction of the lepton spin $\vec\sigma$
with the deuteron $P\slashT$ form factor. 
Effects of a TQFF on polarization observables in
lepton-deuteron scattering have been investigated \cite{edscatt}.
There should be similar effects in proton-deuteron scattering
such as in the planned TRIC experiment at COSY \cite{COSY},
but there they are likely swamped by non-electromagnetic interactions.

We work in the framework of chiral effective field theory (EFT) and take into
account the dominant parity and time-reversal violation in and beyond
the Standard Model (SM) of particle physics. $P$ violation is commonplace
in the weak interaction of the SM. $T$ violation, on the other hand, is small
in the SM, which opens up the possibility that operators involving the SM
fields but having dimension larger than four could be noticeable.
$T$ violation from the CKM quark-mixing matrix is suppressed with respect
to other aspects of weak interactions by a small combination of matrix
elements \cite{Jarlskog}, $J_{C\!P}\simeq 3 \cdot 10^{-5}$.
Moreover, it is loop suppressed in flavor-conserving quantities,
such as $T$-violating form factors of the nucleon and nuclei.
This leaves the QCD vacuum angle $\bar\theta$ \cite{'tHooft}
as the 
potentially largest dimension-four source of such form factors.
However, the stringent experimental limit  on the neutron electric dipole
moment, $|d_n| < 2.9\cdot 10^{-13}\, e$ fm \cite{dnbound},
constrains it to $\bar\theta \simle 10^{-10}$. Therefore, we also
consider $T$ violation originating beyond the SM at a high energy
scale $M_{\slashT}$. The dominant such higher-dimensional
$T$-violating operators are of effective dimension six.

The TQFF is in principle sensitive to $P\slashT$ physics beyond the SM.
However, the lowest dimension where we find $P\slashT$ operators is
eight, which means that, in the simplest
scenarios, they would be highly suppressed by the presumably
high scale of physics beyond the SM. 
Discussions and references on $P\slashT$ interactions at low
energies, including situations where
they could be relatively enhanced, 
can be found in Ref. \cite{PslashT}.
We 
focus here on what is
likely to be the largest ``background'' in the deuteron TQFF: 
the combination of 
$\slashP T$ from the ordinary weak interactions with $\slashPT$ from the 
$\bar\theta$ term and from the dimension-six operators.
Not surprisingly, we find a very small background value for the deuteron
TQFF, so that any experimental evidence for a nonzero TQFF likely results
from new $P\slashT$ physics.

Our discussion is organized as follows. In Section \ref{interactions} 
we construct
the effective chiral Lagrangian for the relevant $PT$, $\slashP T$, 
and $\slashPT$
interactions and currents involving nucleons, pions, and photons. 
In Section \ref{calculation} we calculate
the long-range contributions of these interactions to the deuteron TQFF.
In Section \ref{discussion} we 
discuss 
our results and compare the deuteron TQM
to its $\slashPT$ electric dipole moment (EDM) and magnetic quadrupole moment
(MQM). 
Three Appendices are devoted to details of our calculations.
In Appendix \ref{app:dimsix} the various $\slashPT$ operators are presented 
in more detail, and the orders of magnitude of their contributions are given 
in Appendix \ref{app:ordmag}.
In Appendix \ref{app:ffs} we give the expansion of loop diagrams that define 
the deuteron TQFF.

\section{The effective chiral Lagrangian}
\label{interactions}

At a momentum $Q$ much below the characteristic QCD scale, 
$\MQCD \sim 1$ GeV,
electromagnetic form factors can be calculated with low-energy effective field
theories (EFTs).
The most predictive such an EFT is chiral EFT (for a review,
see Ref. \cite{paulo}), a generalization to an arbitrary
number of nucleons of chiral perturbation theory (ChPT) 
(for a review, see Ref. \cite{veronique}),
where $Q\sim m_\pi$, with $m_\pi\simeq 140$ MeV the pion mass.
In this EFT pion propagation is included explicitly, and the properties
and interactions of the pions are strongly constrained by the approximate
chiral symmetry of QCD. 

For the nucleon, form factors can be calculated 
in perturbation theory as a systematic expansion in $Q/\MQCD$
\cite{veronique}. The $\slashP T$ anapole and the  
$\slashPT$ electric dipole form factor of
the nucleon have been calculated to next-to-leading order (NLO) 
in Refs. \cite{maekawa} and  \cite{BiraHockings},
respectively.
In nuclei, pions can still be treated in perturbation theory \cite{ksw},
but then the expansion is in powers of $Q/\MNN$,
where $\MNN\equiv 4\pi F_\pi^2/m_N\sim F_\pi$ in terms of the 
pion decay constant $F_\pi\simeq 186$ MeV.
For observables involving momenta above $\MNN$,
one-pion exchange needs to be iterated to all orders
\cite{fms}, which complicates renormalization \cite{nogga}.
However, light nuclei are dilute systems and, unless one is interested
in form factors at high momentum, one can use a chiral EFT with 
perturbative pions.
Indeed, the $C$-even electromagnetic form factors of the deuteron, both $PT$
(charge, electric quadrupole, and magnetic dipole) \cite{deutEMFF} and
$\slashPT$ (electric dipole and magnetic quadrupole) \cite{Vri11b}
have been successfully derived in this EFT. 
The 
TDFF 
of the deuteron  has been calculated 
at LO in Ref. \cite{springer}.
Similar calculations could
be performed for other light nuclei.

The relevant low-energy EFT can be written in terms of nucleon, pion,
and photon fields.
The nucleon field $N=(p \; n)^T$ is an isospinor bi-spinor,
with isospin $\boldtau/2$ and spin $S^\mu=(0, {\vec \sigma}/2)$ 
in the rest frame, where the velocity is $v^\mu=(1, {\vec 0})$.
The pion field $\boldpi$ is an isovector pseudoscalar,
for which we choose a stereographic parametrization 
(see, {\it e.g.}, Ref. \cite {Weinberg}) of
the coset space $SO(4)$/$SO(3)$, where $SU(2)\times SU(2)\sim SO(4)$
is the spontaneously broken, approximate chiral symmetry of QCD,
and $SU(2)\sim SO(3)$ its unbroken isospin subgroup.
We define $D\equiv 1+\boldpi^2/F_\pi^2$.
The photon field $A_\mu$ ensures electromagnetic $U(1)$ gauge invariance,
appearing in the gauge and chiral covariant derivatives
$D_\mu \pi_a =D^{-1}
(\delta_{ab}\partial_\mu +e\epsilon_{3ab}A_\mu)\pi_b$
and 
${\cal D}_\mu N=[\partial_\mu +ieA_\mu (1+\tau_3)/2
+ i \boldtau\cdot (\boldpi \times D_\mu\boldpi)/F_\pi^2]N$,
and 
in the field strength $F_{\mu\nu}=\partial_\mu A_\nu - \partial_\nu A_\mu$.
We use the notation $\mathcal D_{\perp\,\pm}^\mu \equiv
\mathcal D_\perp^\mu\pm \mathcal D_\perp^{\dagger\mu}$, where 
$\mathcal D^{\mu}_{\perp} =\mathcal D^{\mu} - v^{\mu} v \cdot \mathcal D$
and $\bar{N} {\cal D}^{\dagger}_\mu =\overline{{\cal D}_\mu N}$.
The coefficients of interactions constructed
with up to two nucleon fields
are estimated, in the absence of other information from QCD, 
by naive dimensional analysis (NDA) \cite{NDA}.
For multi-nucleon couplings 
the scaling of a coefficient on the various scales depends also
on the number of $S$ waves the operator connects \cite{ksw,paulo}.

In the following we will need only a few terms in the 
leading pion-nucleon-photon
$PT$ chiral Lagrangians, {\it viz.}
\begin{eqnarray}
\mathcal L^{(0)}_{PT} &=& 
\frac{1}{2}D_\mu \boldpi \cdot D^\mu \boldpi
-\frac{m_\pi^2}{2D}\boldpi^2
+i\bar N v \cdot {\cal D} N
-\frac{2g_A}{F_{\pi}} (D_\mu\boldpi) \cdot \bar N S^\mu \boldtau N
\nonumber\\
&&- \frac{1}{2} C_0 \left(  \bar N\!N \, \bar N \! N 
- 4 \bar N S^\mu N \cdot \bar N S_\mu N \right) 
+\ldots,
\label{gALag}
\end{eqnarray}
where $g_A\simeq 1.27$ is the nucleon axial coupling
and $C_0$ a contact two-nucleon parameter,
and
\begin{eqnarray}
\mathcal L^{(1)}_{PT} &=& -\frac{1}{2m_N}
\bar N {\cal D}_{\perp}^2N
\nonumber\\
&&
+\frac{e}{4m_N}\epsilon_{\rho\sigma\mu\nu}F^{\rho\sigma} v^\mu 
\bar N \left\{1+\kappa_0
+ (1+\kappa_1) \left[\tau_3
-\frac{2}{F_\pi^2D}
\left(\boldpi^2\tau_3-\pi_3\boldpi\cdot\boldtau\right)\right]\right\} 
S^\nu N 
+\ldots,
\nonumber\\
\label{subPTLag}
\end{eqnarray}
where $\kappa_0\simeq -0.12$ and $\kappa_1\simeq 3.7$ are, respectively, the
isoscalar and isovector anomalous magnetic moments of the nucleon, and 
$\epsilon^{0123}=1$.

The $P\slashT$ TQFF 
vanishes
unless there is, in the EFT, either a $P\slashT$ interaction or a 
combination of $\slashP T$ and $\slashPT$
interactions between the two nucleons. 
$P\slashT$ operators in the EFT Lagrangian arise in two ways.  
First, they represent 
dimension-seven $P\slashT$ operators in the quark-gluon Lagrangian 
just above $\MQCD$. 
These dimension-seven operators in turn can have two origins
above the electroweak scale $v$.
On one hand,
they can be generated by possible gauge-invariant dimension-eight
$P\slashT$ operators,
in which case they would be expected to be suppressed by four powers of 
the high, new-physics scale $M_{\slashT}$,
that is, they would scale as $v/M^4_{\slashT}$.
On the other hand,
they can arise from
the interplay of  
$\slashP T$ in the SM and 
possible dimension-six $\slashPT$ operators,
when one would expect the suppression scale  
to be
$v^2 M^2_{\slashT}$ 
rather than $M^4_{\slashT}$.
A second way to generate $P\slashT$ operators in the EFT Lagrangian 
is from $\slashP T$ and $\slashPT$ interactions
in the quark-gluon Lagrangian at low energy,
when we integrate out non-perturbative dynamics on scale of
order of the typical hadronic scale $\MQCD$.
Again here we expect a suppression of $v^2 M^2_{\slashT}$  rather 
than $M^4_{\slashT}$.

If the new-physics scale is much higher than the electroweak scale, the
contributions from $\slashP T$ and $\slashPT$ interactions
are likely to dominate the $P\slashT$ interactions in the EFT. 
Interesting scenarios in which this is not 
the case are discussed in Ref. \cite{PslashT}.
Here we are interested in the background to genuine $P\slashT$ interactions
at the high energy scale.
In this case, as discussed in App. \ref{app:ordmag},
the contributions from $P\slashT$ interactions in the EFT
are likely smaller than the long-range components from $\slashP T$ 
and $\slashPT$ interactions, which we can, and will, calculate.

$\slashP T$ interactions in chiral EFT have been discussed for example in Refs. 
\cite{Kaplan:1992vj,Zhu:2004vw,maekawa}.
They originate at the QCD scale from four-quark interactions
proportional to the Fermi constant $G_F\simeq 1.2 \cdot 10^{-5}$ GeV$^{-2}$.
A dimensionless measure of the relative strength of the weak interactions
at low energies is $G_FF_\pi^2\sim 4\cdot 10^{-7}$.
The most important interaction is the $\slashP T$ pion-nucleon interaction
\begin{equation}
\mathcal L^{(-1)}_{\slashP T} = \frac{h_1}{F_{\pi}} 
\bar N \left(\boldpi \times \boldtau\right)_3 N +\ldots, 
\label{Parity1}
\end{equation}
with
$h_1 = {\mathcal O}(G_F F^2_{\pi} \MQCD) $.
The $\slashP T$ pion-nucleon coupling $h_1$ is not well-known.
In LO of the EFT with perturbative pions, which we are employing,
the $\slashP T$ asymmetry in $n+p\to d+\gamma$
is $A_\gamma=0.24 h_1/F_\pi$ \cite{npdgammath},
so the recent experimental result 
$A_\gamma=[-1.2 \pm 2.1 (\mathrm{stat}) \pm 0.2 (\mathrm{sys})]\cdot 10^{-7}$ 
\cite{npdgammaexp}
gives a bound 
$| h_1|/F_{\pi} \simle 10^{-6}$,
which is the order of magnitude expected by NDA.
A first lattice QCD calculation at a pion mass $m_\pi\simeq 389$ MeV
gives, in our convention for $h_1$,
$\sqrt{2} h_1/F_{\pi} = [1.099\pm 0.505 ^{+0.058}_{-0.064}]\cdot 10^{-7}$
\cite{hpilatt}.

$\slashPT$ interactions are expected to be due, mostly, 
to the dimension-four QCD $\bar\theta$ term,
parameterized by $\bar\theta \ll 1$, and
the dimension-six operators that result from integrating out physics at the 
scale
$M_{\slashT}$ and the heavy degrees of freedom in the SM. The complete set of
$\slashPT$ dimension-six operators at the electro-weak scale has been given 
in Ref. 
\cite{Buchmuller:1985jz}, and the relevant operators at the hadronic scale 
have been
summarized in Ref. \cite{deVries:2012ab}.
They are the isoscalar and isovector quark EDM (qEDM) and quark chromo-EDM
(qCEDM), the Weinberg operator, which gives rise to a gluon chromo-EDM (gCEDM),
and four $\slashPT$ four-quark operators. Two of these four-quark operators
are invariant under the SM gauge group and can be generated directly at the
electroweak scale. Their effect in the chiral EFT at low energy cannot be 
separated
from the gCEDM and we refer to these collectively as chiral-invariant sources 
($\chi$ISs).
The other two four-quark operators break isospin
and result from integrating out the weak gauge bosons and running to low energy.
Because they mix left- and right-handed quarks we denote these as FQLR.
The various $\slashPT$ sources are further discussed in 
App. \ref{app:dimsix}.

The dimension-four and six $\slashPT$ operators have different transformation
properties under the chiral group $SU_L(2) \times SU_R(2)$, which has 
consequences
for the $\slashPT$ couplings in  chiral EFT \cite{BiraEmanuele,deVries:2012ab}. 
The interactions
relevant to the 
rest of the paper are
\begin{eqnarray}
\mathcal L_{\slashPT} &=& 
-\frac{\bar g_0}{F_{\pi}} \bar N \boldpi\cdot\boldtau N 
-\frac{\bar g_1}{F_{\pi}} \pi_3 \bar N N 
-2 \bar N \left(\bar d_0 + \bar d_1 \tau_3 \right) S^{\mu} 
\left( v^{\nu}+ \frac{i \mathcal D^{\nu}_{\perp\, -}}{2 m_N}\right) N F_{\mu \nu}
\nonumber \\
& &+ \frac{1}{4} \bar C_0 \left[  \bar N\!N \, \partial_{\mu} (\bar N S^{\mu} N )
- \bar N \boldtau N \cdot \mathcal D_{\mu} 
\left( \bar N S^{\mu} \boldtau N \right) \right] ,
\label{g0Lag}
\end{eqnarray}
where $\bar g_0$ ($\bar g_1$) is the isoscalar (isovector)
$\slashPT$ pion-nucleon coupling,
$\bar d_0$ ($\bar d_1$) a short-range contribution to the isoscalar (isovector)
nucleon EDM, and 
$\bar C_0 $ a short-range $\slashPT$ two-nucleon interaction. 
The term
proportional to $1/m_N$ is a recoil correction and depends on the sum of the 
incoming and outgoing nucleon momenta. 
Other $\slashPT$ interactions, some expected to be of comparable size, will not
be needed below because of the quantum numbers of the deuteron.

The relative importance of the operators in Eq. \eqref{g0Lag} depends on the 
chiral properties
of the $\slashPT$ source at the quark-gluon level. 
As described in  App. \ref {app:ordmag},
the dimensionless one-nucleon couplings $\bar g_{0,1}/\MQCD$ and 
$\MQCD \bar d_{0,1}/e$
are given by the dimensionless strengths of the underlying $\slashPT$ 
interactions, times factors of $(m_\pi/\MQCD)^2$ that depend on the chiral
transformation properties of the source.
For the QCD $\bar\theta$ term, the qCEDM, and the FQLR, which violate chiral 
symmetry,
non-derivative pion-nucleon couplings like $\bar g_0$ 
can appear in the chiral 
Lagrangian at LO.
In this case $\bar g_0 /\MQCD = {\cal O}(\MQCD\bar d_1/e)$
and pion effects tend to dominate because of the low mass.
In contrast, 
$\chi$ISs can generate pion-nucleon non-derivative couplings  
only through insertion of the quark mass, 
which costs two powers of $m_\pi/\MQCD$,
so that, for example,  
$\bar g_0 /\MQCD = {\cal O}( (m_\pi/ \MQCD)^2 \MQCD\bar d_1/e)$.
The $\bar g_0$ term still appears in the LO Lagrangian, 
but it is accompanied by the
equally important two-nucleon and electromagnetic operators, 
whose
construction does not require any insertion of the quark mass. 
Finally, the presence of a photon
field causes the qEDM to contribute mainly to 
the photon-nucleon sector, 
purely hadronic operators being suppressed by powers of 
$\alpha_{\textrm{em}}/4\pi$.
 
The interactions in Eqs. \eqref{Parity1} and \eqref{g0Lag} can be used to 
compute the
$\slashP T$, $\slashPT$ and $P \slashT$ form factors of nuclei.
The nucleon does not possess a $P\slashT$ form factor. 
We summarize here the results
for the nucleon TDFF and electric dipole form factor (EDFF), 
which are needed for the calculation of the deuteron
TQFF in Sec. \ref{calculation}.
The $\slashP T$  and $\slashPT$ currents are written as, respectively,
\begin{equation}
J^{\mu}_{\slashPsub T}(q) = \frac{2}{m_N^2} 
\left( F_{\slashPsub T,\, 0}(-q^2) + F_{\slashPsub T, 1}(-q^2) \tau_3\right) 
\left[S^{\mu} q^2 - S \cdot q q^{\mu} +\ldots\right]  
\label{currentTDFF}
\end{equation}
and
\begin{equation}
J^{\mu}_{\slashPTsub}\left(q,K\right) = 2 i
\left( F_{\slashPTsub,\, 0}(-q^2) +  F_{\slashPTsub,\, 1}(-q^2) \tau_3\right) 
\left[S^{\mu}\left( v\cdot q +\frac{K\cdot q}{m_N}\right) - S\cdot q \left(v^{\mu}
+ \frac{K^{\mu}}{m_N} \right)  +\ldots\right],
\label{currentEDFF}
\end{equation}
where $q$ denotes the four-momentum of the photon and $2 K$ is the sum of
the nucleon momenta.
We write
\begin{equation}
F_{\slashPsub T,\, i}(-q^2) = a_i \ f_{ i}\left(-q^2/4m_\pi^2\right),
\end{equation}
and
\begin{equation}
F_{\slashPTsub,\, i}(-q^2) = d_i 
-q^{2} \ S^{\prime}_{i}\left(-q^2/4m_\pi^2\right) ,
\end{equation}
where $a_0$ and $a_1$ ($d_0$ and $d_1$) are the nucleon isoscalar and isovector 
anapole (electric dipole) moments,
$f_{i}(0)$=1, and $S^{\prime}_{i}(0)$ is finite.

At LO, the nucleon TDFFs
come entirely from pion loops, in which one vertex
is the $\slashP T$ pion-nucleon coupling $h_1$.
By NDA one expects 
$a_i/m_N^2= {\cal O} (e h_1/m_\pi \MQCD^2)$.
The calculation of  
Ref. \cite{maekawa} shows that the nucleon anapole form factor is, 
at LO, isoscalar and finite,
\begin{equation}
a_0^{({\rm LO})} = \frac{e g_A h_1 m_N^2}{24\pi F^2_{\pi} m_{\pi}}, 
\qquad 
f_{0}^{({\rm LO})}\left(x^2\right)= \frac{3}{2x^2}
\left[\frac{1+x^2}{x}\arctan x-1\right] ,
\qquad  
a_1^{({\rm LO})} = 0.
\label{LeadingAnapolescalar}
\end{equation}
The isovector anapole form factor appears only at NLO,
where short-range contributions to the moments also are present.
Neglecting ${\cal O}(1)$ numbers, the result \eqref{LeadingAnapolescalar} 
for $a_0^{({\rm LO})}$ is
a factor of $4\pi$ larger than the NDA estimate, as often happens in 
baryon ChPT.

The nucleon EDFF was computed in Ref. \cite{BiraHockings}
to NLO for all $\slashPT$ sources of dimension up to six.
For the QCD $\bar\theta$ term, the qCEDM, and the FQLR, the isovector 
nucleon EDM 
receives a one-loop contribution from $\bar g_0$ at LO.
At the same order there are also 
short-range isoscalar ($\bar d_0$) and isovector ($\bar d_1$) 
contributions, 
the latter being 
required by renormalization-group invariance. 
The isoscalar and isovector
nucleon 
EDMs are given by
\cite{CDVW79}
\begin{equation}
d_0^{({\rm LO})} = \bar d_0, 
\qquad
d_1^{({\rm LO})} = \bar d_1 (\mu)
+ \frac{e g_A \bar g_0}{\left(2\pi F_{\pi}\right)^2} 
\left( L - \ln \frac{m^2_{\pi}}{\mu^2}\right), 
\label{LeadingEDM}
\end{equation}
where we used dimensional regularization
in $d$ spacetime dimensions, 
with $L=2/(4-d)-\gamma_E+\ln 4\pi$,
and $\mu$ the renormalization scale.
In this case there is no $4\pi$ enhancement, and 
the nucleon EDM is suppressed by the loop
factor $(2\pi F_{\pi})^2 \sim \MQCD^2$ with respect to the pion nucleon coupling
$\bar g_0$.
The momentum dependence of the EDFF
is purely isovector in LO 
and governed by the scale $m_{\pi}$, 
as is the case for the isoscalar TDFF \eqref{LeadingAnapolescalar},
but it is not needed in the following. 
For the qEDM and the $\chi$ISs, $e\bar g_0/m_\pi^2$ is at most as large as the 
short-range coupling
$\bar d_{1}$, 
and the loop suppression makes its contribution negligible. 
The EDFF is then momentum independent at LO 
and completely determined by the low-energy constants
$\bar d_{0,1}$,
\begin{equation}
d_{0,1}^{({\rm LO})} = \bar d_{0,1}, 
\qquad 
S^{\prime\, \textrm{(LO)}}_{0,1}\left(x^2\right)= 0.
\label{LeadingEDMprime}
\end{equation}
In this case the momentum dependence appears in higher order and
is determined by short-range physics.

The 
$\slashPT$ couplings 
$\bar g_0$, $d_1$, and $\bar C_0$ are not known
and,
in order to estimate the magnitude of the TQFF they induce,
we will need to make some reasonable assumptions.
First, we assume that there are no cancellations between $d_0^{({\rm LO})}$
and $d_1^{({\rm LO})}$, so that,
for $\slashPT$ violation from 
the qEDM and $\chi$ISs, the bound on the neutron EDM $|d_n|$
can be directly translated into the bound 
$| \bar d_1 |< 2.9 \cdot 10^{-13}\, e$ fm. 
Second, as pointed out in Ref. \cite{CDVW79}, we should not expect any
cancellation  in Eq. \eqref{LeadingEDM} between pieces
that are non-analytic and analytic in $m_\pi^2$.
With the reasonable value $\mu = m_N$, the same bound 
applies for $| \bar d_1 (m_N)|$ 
in the case of $\bar\theta$ term, qCEDM and FQLR.
Moreover, since the long-range contributions give
the estimate 
$|d_1|\sim 0.13 (|\bar g_0|/F_{\pi}) \, e$  fm,
the existing experimental bound on the neutron EDM 
yields an approximate bound on the $\slashPT$ pion-nucleon coupling,
$|\bar g_0|/F_{\pi} \simle 2\cdot 10^{-12}$.

\section{TQFF of the deuteron}
\label{calculation}
 
With the interactions described in Sec. \ref{interactions}, we can calculate 
the long-range
contributions to the deuteron TQFF,
using the techniques of Refs. \cite{deutEMFF,springer,Vri11b}.
As usual in such a calculation, the orders of magnitude of the various 
contributions can be found by combining
the power counting rules of ChPT based on NDA
with the rules for two-nucleon states as
summarized, for example, in Ref. \cite{paulo}.
A pion propagator scales as $1/Q^2$.
A loop involving a single nucleon contributes a factor $Q^4/(4\pi)^2$
from the integration and a factor $1/Q$ from the nucleon propagator.
The infrared enhancement of a loop involving two nucleons gives a
factor $Q^5/4\pi m_N$ from the integration and 
a factor $m_N/Q^2$ from each nucleon propagator.
The deuteron wavefunction contributes an overall normalization
factor $4\pi Q/m_N^2$.

The deuteron itself is built out of the two-nucleon contact interaction
with coefficient $C_0={\mathcal O}(4\pi/m_N \gamma)$ and the nucleon kinetic 
terms in Eqs. \eqref{gALag} and \eqref{subPTLag}.
Pion exchange originating from the pion kinetic terms and
pion-nucleon coupling in Eq. \eqref{gALag} contributes to the deuteron structure
at relative ${\cal O}(Q/\MNN)$, together with a two-derivative
contact interaction that accounts for short-range energy dependence in the
on-shell two-nucleon amplitude \cite{ksw}.
Since we calculate the TQFF to LO only, $\gamma$ is the sole $PT$ two-nucleon
input needed.

\begin{figure}[t]
\center
\includegraphics[width=15cm]{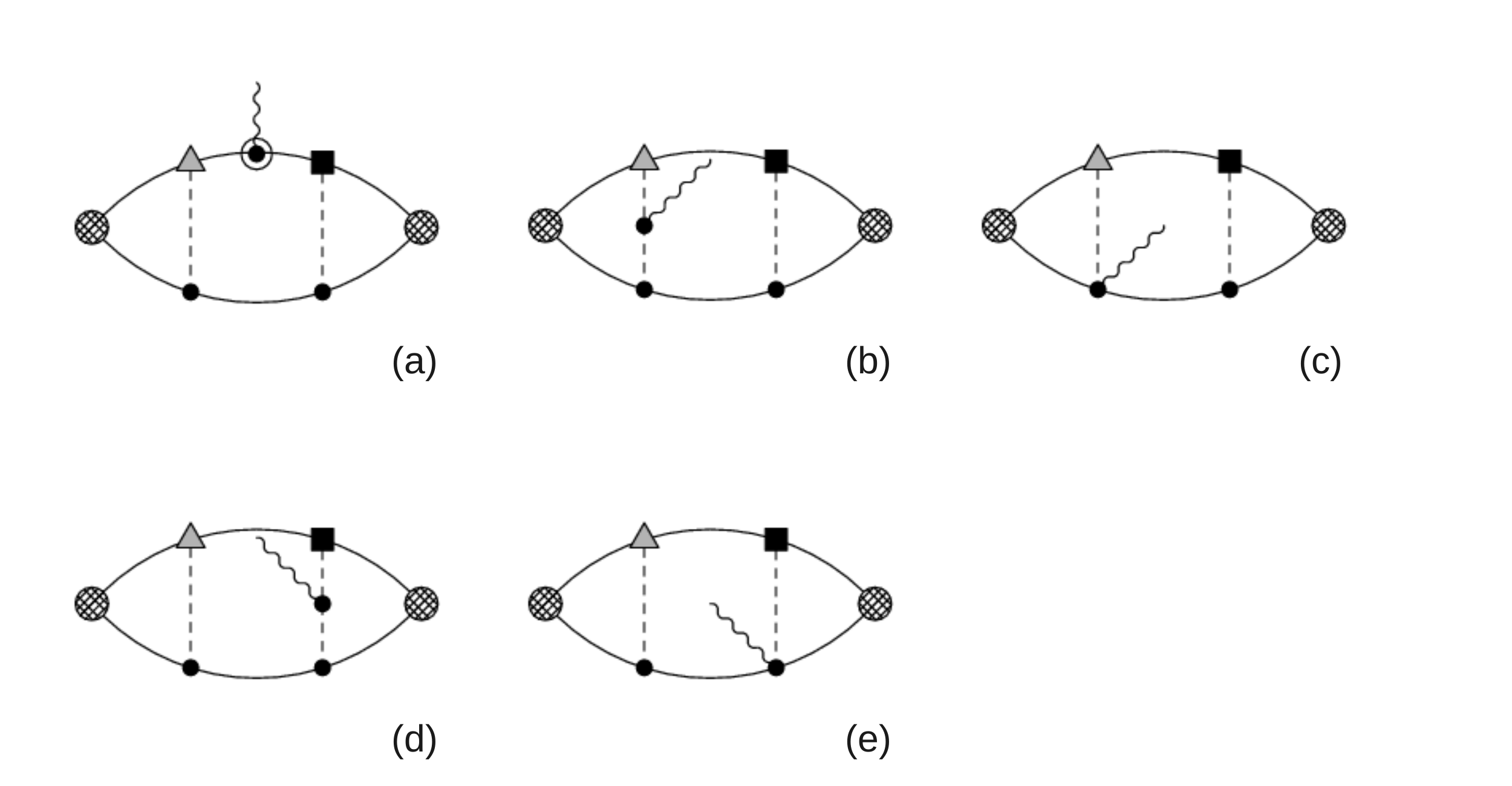}
\caption{Two-pion-exchange (TPE) contributions to the deuteron TQFF, 
$F_{P\slashT}(\vec q^{\, 2})$.
Nucleons. pions and photons are represented by solid, dashed and wavy lines, 
respectively.
LO $PT$, $\slashPT$, and $\slashP T$ interactions are denoted by circles, 
squares, and
triangles, respectively. 
An NLO $PT$ interaction is denoted by a circled circle.
Deuteron states obtained from the
iteration of the leading $PT$ two-nucleon interaction are represented 
by hatched circles.}
\label{Fig1}
\end{figure}

\begin{figure}[t]
\center
\includegraphics[width=15cm]{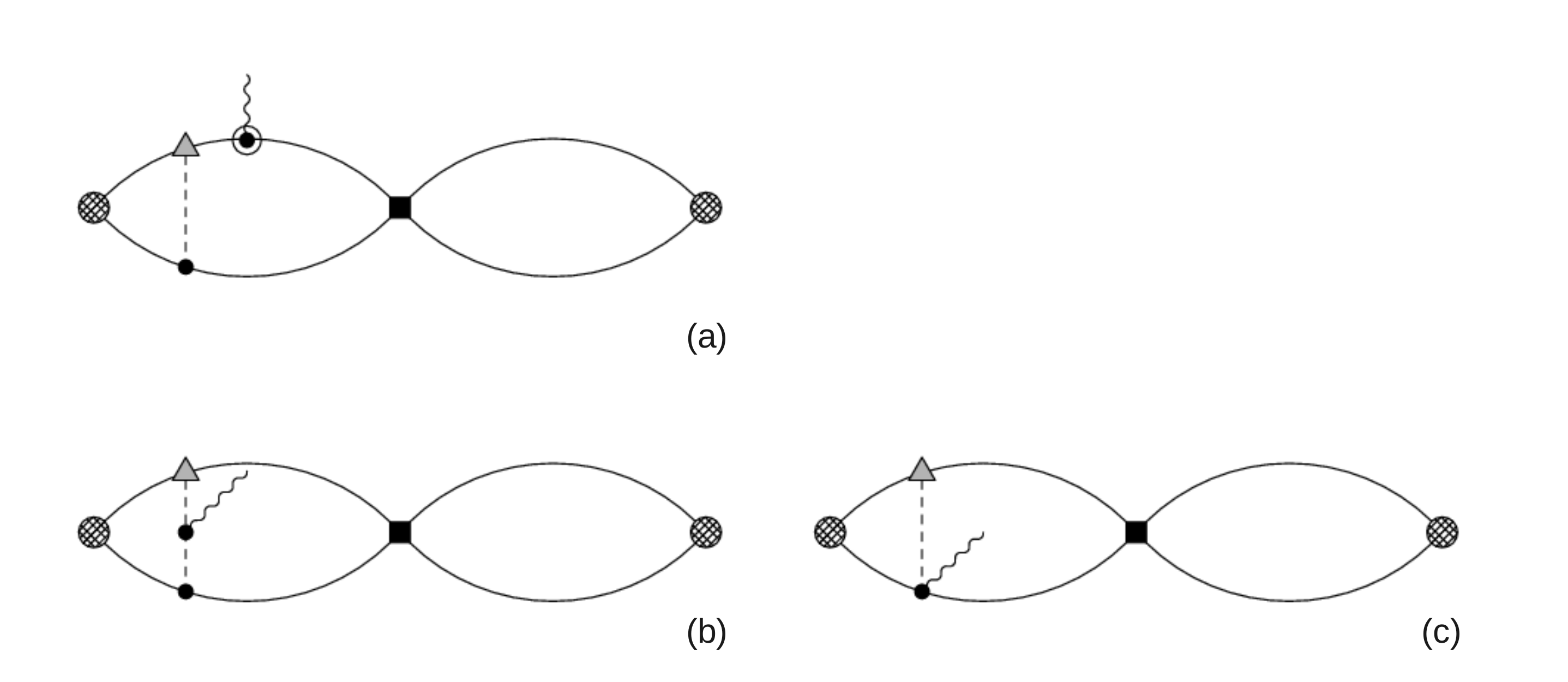}
\caption{Short-range two-nucleon ($4N$) contributions to the deuteron TQFF, 
$F_{P\slashT}(\vec q^{\, 2})$. The notation is as in Fig. \ref{Fig1}.}
\label{Fig2}
\end{figure}

\begin{figure}
\center
\includegraphics[width=15cm]{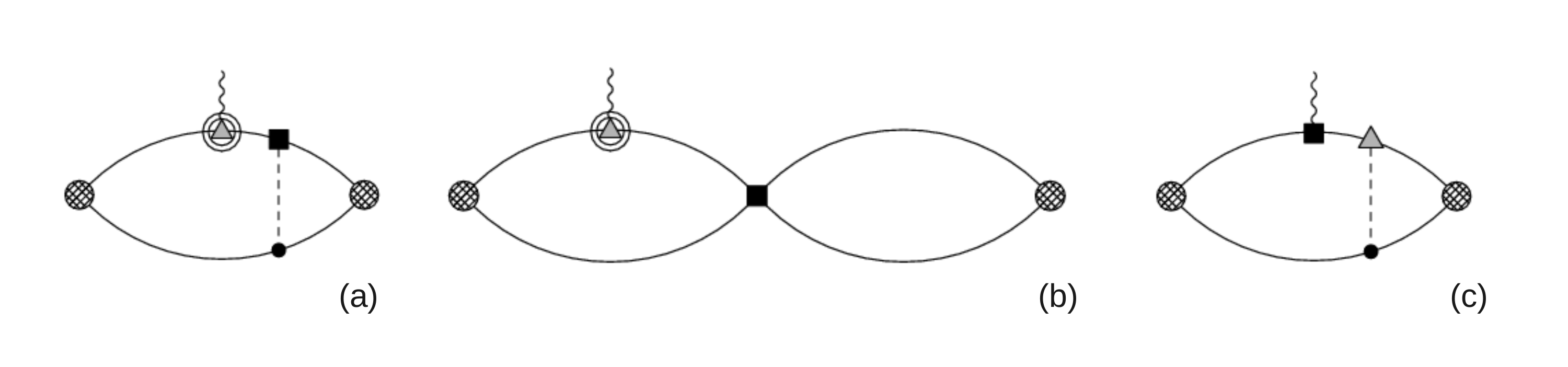}
\caption{Nucleon anapole form factor (TDFF)
and electric dipole moment (EDM) contributions to the 
deuteron TQFF, $F_{P\slashT}(\vec q^{\, 2})$. 
The twice-circled triangle stands for the anapole form factor. 
The other notation is as in Fig. \ref{Fig1}.}
\label{Fig3}
\end{figure}

The $P\slashT$ TQFF is an intrinsically two-nucleon observable,
which requires at least one symmetry-violating interaction between
the two nucleons.
We argue in App. \ref{app:ordmag} that 
$P \slashT$ interactions are much smaller than contributions
from separate $\slashP T$ and $\slashPT$ interactions.
The lowest-order diagrams involving the $\slashP T$ vertex $h_1$ 
and one of the $\slashPT$
couplings 
are shown in Figs. \ref{Fig1}--\ref{Fig3}.
In these figures, only one possible ordering is shown.
Circles, triangles, and squares denote the leading $PT$, $\slashP T$, and 
$\slashPT$
interactions 
in Eqs. \eqref{gALag}, \eqref{Parity1}, and \eqref{g0Lag}, respectively;
a circled circle, the $PT$ magnetic photon-nucleon interactions
in Eq. \eqref{subPTLag};
a twice circled triangle, the $\slashP T$ anapole moment of the nucleon
in Eq. \eqref{LeadingAnapolescalar}.
The hatched circles denote deuteron states obtained from the iteration of the
leading two-nucleon interaction,
which brings in dependence on the binding momentum $\gamma$. 
The natural scale for momentum dependence of the TQFF
is $4\gamma$, so we express our results in terms of $\vec x = \vec q/4\gamma$.
We also define the ratio $\xi = \gamma/m_{\pi}$ of low-momentum scales.

Let us first consider a photon which interacts without breaking $P$ and $T$. 
In this case the photon couples to the nucleon via the
magnetic couplings in Eq. \eqref{subPTLag}
or to a pion via
interactions obtained by gauging the derivatives in
the pion kinetic energy and pion-nucleon axial coupling in Eq. \eqref{gALag}.
Diagrams with only one pion exchange and $\bar g_0$ and $h_1$ vertices on each 
end vanish.
This can be  
understood from the fact that such diagrams do not 
have enough powers of
momentum in the vertices to generate a form factor of the form of 
Eq. \eqref{PCTVFF}, and it agrees with 
the more general analysis of the $P\slashT$ two-nucleon interaction 
\cite{simonius}.
This leaves three-loop diagrams, containing either two pion exchanges (TPE)
or one pion exchange 
and a short-range $\slashPT$ two-nucleon interaction (4N),
Figs. \ref{Fig1} and \ref{Fig2} respectively.
Using the power-counting rules outlined above, 
the sizes of the diagrams in Figs. \ref{Fig1} and \ref{Fig2} are
\begin{eqnarray}
\textrm{Fig. \ref{Fig1}} &=& 
\mathcal O\left(\frac{e h_1}{Q^2 \MNN^2} 
\frac{\bar g_0}{\MQCD}\right), 
\label{Fig1scaleprime}\\
\textrm{Fig. \ref{Fig2}} &=& 
\mathcal O\left(\frac{e h_1}{Q^2 \MNN^2} 
\frac{ m_N \gamma \bar C_0}{4\pi}\frac{Q \MNN}{\MQCD}\right).
\label{Fig2scaleprime}
\end{eqnarray}
Whether the diagrams in Figs. \ref{Fig1} or \ref{Fig2} are more important 
depends on the $\slashPT$ source.
For the $\tb$ term, the qEDM, the qCEDM, and the FQLR operator 
the contributions from the short-range interaction $\bar C_0$ are always 
suppressed, in this case by $Q/\MNN$,
with respect to TPE, because for these sources 
$\bar g_0 =\mathcal O(\MNN^2 m_N \gamma \bar C_0/4\pi)$, 
see App. \ref{app:ordmag}.
For $\chi$ISs, the opposite is true
because of the extra $(Q/\MQCD)^2$ suppression of $\bar g_0/\MQCD$,
which makes the short-range contributions larger by 
a factor of $\mathcal O(\MNN/Q)$.
The diagrams in Fig. \ref{Fig1} are formally the leading contributions 
for the QCD $\bar \theta$
term, the qCEDM, and the FQLR, while those in Fig. \ref{Fig2} are leading 
for the $\chi$ISs. 
Note that for the isovector qCEDM and the FQLR, one should consider
not only the isoscalar pion-nucleon coupling $\bar g_0$
but also the isovector pion-nucleon coupling $\bar g_1$,
but such diagrams vanish. 

Alternatively, the photon can interact with the nucleon with a $\slashP T$
or $\slashPT$ interaction, in which case a single two-nucleon
interaction, $\slashPT$ or $\slashP T$ respectively, is sufficient
to produce a TQFF --- see Fig. \ref{Fig3}.
In diagrams \ref{Fig3}(a,b) one of the nucleons couples to the magnetic 
field via
its anapole moment, with $\slashPT$ coming either from  pion exchange or 
from a two-nucleon interaction. Here the anapole ``vertex''
stands for a one-loop diagram, which produces the result 
\eqref{LeadingAnapolescalar}. 
In diagram \ref{Fig3}(c) the photon couples to the nucleon through 
a recoil correction to the
$\slashPT$ EDM, with $\slashP T$ coming from pion exchange.
By power counting, the contributions of diagrams \ref{Fig3}(a,b) to the TQFF 
are
\begin{eqnarray}
\textrm{Fig. \ref{Fig3}(a)} &=& 
\mathcal O\left( \frac{a_i}{m_N^2} \frac{\bar g_0}{Q \MNN} \right)
 = \mathcal O\left(\frac{e h_1}{Q^2 \MNN^2}
\frac{\bar g_0 \MNN}{\MQCD^2}\right), 
\label{Fig3ascaleprime}\\
\textrm{Fig. \ref{Fig3}(b)} &=& 
\mathcal O\left( \frac{a_i}{m_N^2} \frac{m_N \g \bar C_0 }{4\pi }\right) 
= \mathcal O\left( \frac{eh_1}{Q^2 \MNN^2} 
\frac{m_N \gamma\bar C_0 }{4\pi } \frac{Q\MNN^2}{\MQCD^2} \right) 
\label{Fig3bscaleprime},
\end{eqnarray}
were we used the NDA expectation for the anapole moment.
Diagrams \ref{Fig3}(a,b) are thus
suppressed by one power of $\MNN/\MQCD\sim 1/4\pi$
compared to the contributions of Figs. \ref{Fig1} and \ref{Fig2}.
However, $a_0$ in  
Eq.~\eqref{LeadingAnapolescalar} is 
a factor of $4\pi$ larger than the NDA estimate,
making the corresponding contributions to the TQFF competitive with LO.
Again, of other possible $\slashPT$ 
couplings only $\bar g_0$ and $\bar C_0$ contribute.
Diagram  3(c) represents contributions to the TQFF coming from the nucleon EDFF.
It scales as
\begin{equation}\label{Fig3cscaleprime}
\textrm{Fig. \ref{Fig3}(c)} = 
\mathcal O\left(  \frac{eh_1}{Q^2\MNN^2} \frac{ \bar d_1}{e} 
\frac{Q\MNN}{\MQCD} \right),
\end{equation}
and there is no contribution from $\bar d_0$ at this order.
For $\bar\theta$ term, qCEDM, and FQLR, 
$\bar d_1 /e=\Or(\bar g_0/\MQCD^2)$ and this contribution
is suppressed by $Q/\MQCD$ (a factor coming from the recoil)
compared to the analogous anapole diagram \ref{Fig3}(a).
For $\chi$ISs, this contribution is comparable to Fig. \ref{Fig2},
while for the qEDM it is the sole leading contribution,
since numerically $\alpha_{\mathrm{em}}/4\pi\sim (Q/\MQCD)^3$.

To summarize these power-counting arguments, we expect the TQFF induced by
the $\bar\theta$ term, the qCEDM,  and the FQLR to be dominated by the TPE
diagrams in Fig. \ref{Fig1}, with possibly large corrections from 
the nucleon TDFF,
diagram \ref{Fig3}(a). For $\chi$ISs, the dominant contribution should come
from the diagrams involving the $\slashPT$ short-range two-nucleon
interaction in Fig. \ref{Fig2} 
and from the nucleon EDFF, diagram  \ref{Fig3}(c), 
with a sizable correction from the nucleon TDFF, diagram \ref{Fig3}(b). 
For the qEDM only the nucleon EDFF contribution \ref{Fig3}(c)
should be important.

We now proceed to the evaluation of diagrams in Figs. \ref{Fig1},
\ref{Fig2}, and \ref{Fig3}.
We find that only the isovector magnetic moment gives a non-vanishing 
contribution to diagram \ref{Fig1}(a) and \ref{Fig2}(a).  
Diagrams with other photon-nucleon interactions  
(the isoscalar magnetic
moment and the minimal coupling through the covariant derivative 
in the nucleon kinetic term) vanish.
Similar diagrams where pion exchanges and contact interaction
occur both before or after the insertion of the photon coupling also vanish.
Diagrams \ref{Fig1}(b,c) and \ref{Fig1}(d,e) differ only by an isospin factor. 

The diagrams in Fig. \ref{Fig1} are finite in four (and three) dimensions. 
We express the result for their contribution 
to the TQFF as 
\begin{equation}\label{I3ab}
F_{P \slashT}^{(\textrm{TPE})}(\vec q^{\, 2})  = 
 -\frac{e g_A^2 \bar g_0 h_1}{m^2_{\pi}} 
\frac{m_N}{\left(4\pi F^2_{\pi}\right)^2} \left[ (1+\kappa_1)   
\, I^{(3)}_a\left( \frac{\gamma}{m_{\pi}}, \frac{\vec q}{4\gamma}\right)
+  I^{(3)}_b\left(\frac{\gamma}{m_{\pi}}, \frac{\vec q}{4\gamma}\right) 
\right],
\end{equation}
in terms of  two three-loop integrals $I^{(3)}_{a,b}(\xi, \vec x)$.
Likewise,
the results for the TQFF in Fig. \ref{Fig2} are expressed 
in terms of two two-loop functions
$I_{a,b}^{(2)}(\xi, \vec x)$ as
\begin{equation}
F_{P \slashT}^{(\textrm{4N})}(\vec q^{\, 2})  =
\frac{eg_A h_1}{m_{\pi}} \frac{m_N}{4\pi F_{\pi}^2}\frac{\mu - \gamma}{4\pi} 
\bar C_0 
\left[  (1+\kappa_1)   
\, I^{(2)}_a\left( \frac{\gamma}{m_{\pi}}, \frac{\vec q}{4\gamma}\right)
+ I^{(2)}_b\left( \frac{\gamma}{m_{\pi}}, \frac{\vec q}{4\gamma}\right) 
\right],
\label{I2ab}
\end{equation}
where we have used power-divergence subtraction \cite{ksw}.
The $\mu$ dependence is absorbed in $\bar C_0$ itself,
since here it appears in the same combination as in the magnetic quadrupole
form factor \cite{Vri11b}.
The expansions to order $\vec{q}\,^2$
of the integrals $I^{(2,3)}_{a,b}$ 
are given in App. \ref{app:ffs}.
The resulting contributions to the TQM are
\begin{equation}
\mathcal T_d^{({\rm TPE})}\simeq 
\left[  0.8\, (1+\kappa_1) - 0.9  \right] \cdot 10^{-2}
\; \frac{\bar g_0 h_1}{F_{\pi}^2} \; e \, \textrm{fm}^3.
\label{valueLO}
\end{equation} 
and
\begin{equation}\label{LOvalue4N}
\mathcal T^{(\textrm{4N})}_d  \simeq
\left[ 1.0\,  (1+\kappa_1)  - 0.7 \right] \cdot 10^{-2} 
\, m_N \frac{\mu - \gamma}{4\pi} \bar C_0 \, h_1  \, e\, \textrm{fm}^3.
\end{equation}

Finally,
we consider 
diagrams \ref{Fig3}(a,b) 
and (c).
For isoscalar $\slashPT$, only the isoscalar TDFF $F_{\slashPsub T,\, 0}$ 
contributes in diagrams \ref{Fig3}(a,b).
Isospin-breaking $\slashPT$, for example from insertions of $\bar g_1$, 
would contribute to diagram \ref{Fig3}(a)
together with the isovector TDFF $F_{\slashPsub T,\, 1}$.
However, the isovector TDFF is suppressed by $Q/\MQCD$ (and by a factor
of $4\pi$) with respect to the isoscalar TDFF.
Therefore, even for sources that generate $\bar g_0$ and $\bar g_1$ at the
same level, $\bar g_1$ contributions to the TQFF are subleading.
Diagram \ref{Fig3}(c)
is leading only for the qEDM and $\chi$ISs, for which the isovector EDFF 
is momentum independent and coincides with the EDM. 
Diagrams \ref{Fig3}(a,b) and (c)
result in contributions to the TQFF given by
\begin{eqnarray}\label{I2c}
F_{P \slashT}^{({\rm TDFF})}(\vec q^{\, 2}) =
\frac{F_{\slashPsub T,\, 0}(\vec q^{\, 2})}{4\pi m_N} 
\left[ \frac{g_A \bar g_0}{m_{\pi}F^2_{\pi}} 
\, I_c^{(2)}\left(\frac{\gamma}{m_{\pi}},\frac{\vec q}{4\gamma}\right) 
+  (\mu-\gamma) \bar C_0  
\; I^{(1)}\left(\frac{\vec q}{4\gamma}\right)
\right]
\label{FF1}
\end{eqnarray}
and
\begin{equation}\label{nEDM}
F_{P\slashT}^{({\rm EDM})}(\vec q^{\, 2}) = 
\frac{g_A}{4\pi F^2_{\pi} m_{\pi}}  h_{1} \bar d_1 
\; I^{(2)}_d \left(\frac{\gamma}{m_{\pi}}, \frac{\vec q}{4\gamma}\right),
\end{equation}
respectively.
The expression for the one-loop integral $I^{(1)}$ along with
the expansions to order $\vec{q}\,^2$ of two-loop integrals
$I^{(2)}_{c,d}$ can be found in App. \ref{app:ffs}.
Numerically this gives
\begin{equation}
\mathcal T_d^{({\rm TDFF})}  
\simeq \left[ 3.5 \frac{\bar g_0}{F^2_{\pi}} 
+ 2.7 \, m_N \frac{\mu - \gamma}{4\pi } \bar C_0\right] \cdot 10^{-2 }
\; h_1 \; e \, \textrm{fm}^3
\label{valueLO'}
\end{equation}
and
\begin{equation}
\label{TQMEDM}
\mathcal T_d^{({\rm EDM})}
\simeq 1.3 \cdot 10^{-2} \; \bar d_1 h_1 \,  \textrm{fm}^3.
\end{equation}

The result \eqref{valueLO'} shows that the TDFF contribution, 
though expected to be suppressed by the factor $\MNN/\MQCD$, is  
comparable to the LO values in Eqs.~\eqref{valueLO} and \eqref{LOvalue4N},
in line with the $4\pi$ enhancement in the TDM.
For the QCD $\bar\theta$ term, the qCEDM, and the FQLR, we can assess the 
importance
of the EDM contribution to the TQFF by
substituting in Eq. \eqref{TQMEDM} the estimate for the nucleon EDM in terms of
$\bar g_0$, $|d_1| \sim 0.13 (|\bar g_0|/F_{\pi}) \, e$ fm. We find  
$\mathcal T_d^{({\rm EDM})}\sim 0.2\cdot 10^{-2} (\bar g_0 h_1/F^2_{\pi})\, e$ fm$^3$,
which is numerically small compared to Eqs. \eqref{valueLO} 
and \eqref{valueLO'}, as
expected by power counting.

We can now combine the results found so far.
For the QCD $\bar\theta$ term, the qCEDM, and the FQLR, the TPE contributions
from the diagrams in Fig. \ref{Fig1} and the TDFF contributions 
in Fig. \ref{Fig3}(a)
have comparable size, giving
\begin{equation}\label{Tdtheta}
(\mathcal T_d)_{\bar\theta,\, \textrm{qCEDM},\, \textrm{FQLR}}\simeq 
6.3 \cdot 10^{-2} \; \frac{\bar g_0 h_1}{F^2_{\pi}} \, e \,\textrm{fm}^3.
\end{equation} 
This number is within a factor $\simeq 2$ of
the power counting estimate in Eq. \eqref{Fig3ascaleprime},
indicating that the power counting works well (apart from
the 
$4\pi$ in the anapole moment).

For $\chi$ISs, by power counting the  leading contributions are expected to 
come from the diagrams in Fig. \ref{Fig2}, with insertions of the 
four-nucleon coupling $\bar C_0$, and from the EDM
in diagram \ref{Fig3}(c). Also in this case, the contribution of the TDFF
in diagram \ref{Fig3}(b) is numerically important.
We find
\begin{equation}
(\mathcal T_d)_{\textrm{$\chi$ISs}}  \simeq
\left[ 6.7 \, m_N \frac{\mu - \gamma}{4\pi} \bar C_0  
+ 1.3\,  \frac{\bar d_1}{e} \right]\cdot 10^{-2 } \; h_1 \;e \,\textrm{fm}^3 .
\label{TdchiISs}
\end{equation}
If $\bar C_0$ and $\bar d_1$ have their NDA values, 
their respective contributions are numerically comparable.

In the case of the qEDM, the TQM is dominated by the contribution from the 
nucleon EDM, and we find
\begin{equation}\label{TdqEDM}
(\mathcal T_d)_{\textrm{qEDM}} \simeq 1.3 \cdot 10^{-2} \; \bar d_1 h_1 
 \,\textrm{fm}^3.
\end{equation}
Because of dimensionless numerical factors this value is about an order of magnitude smaller
than expected by the power-counting estimates based on NDA.

\section{Discussion and 
conclusion} 
\label{discussion}

It is interesting to compare our results for the deuteron TQM in 
Eqs. \eqref{Tdtheta}, \eqref{TdchiISs},
and \eqref{TdqEDM} with the largest $\slashPT$ moment of the deuteron, 
EDM or MQM, for the
respective $\slashPT$ sources. 
In Ref. \cite{Vri11b} power-counting estimates and
LO results were given for the deuteron EDM, $d_d$, and MQM, ${\mathcal M}_d$, 
in chiral EFT with  perturbative pion exchange.

For sources that break chiral
symmetry and generate non-derivative $\slashPT$ pion-nucleon couplings in LO, 
$d_d$ and ${\mathcal M}_d$ are expected to be dominated by two-body effects
and be enhanced with respect to the nucleon EDM. 
In the case of the QCD $\bar\theta$ term,
${\mathcal M}_d$
is expected to be the largest moment (in natural units), 
because at LO $\bar g_0$ does not contribute
to $d_d$
(except through the nucleon EDM, Eq. \eqref{LeadingEDM}). On the other
hand, the qCEDM and the FQLR, which generate also 
the isovector coupling $\bar g_1$ in LO,
induce $d_d$ and ${\mathcal M}_d$
of the same size.
For $\chi$ISs, 
$d_d$ and ${\mathcal M}_d$ are also expected to be of the same size, 
and of similar
size as the nucleon EDM. 
The deuteron EDM is 
in fact expected to be well approximated
by (twice) 
the isoscalar nucleon EDM, 
while the deuteron MQM, 
in the perturbative-pion counting, receives the largest contribution from 
the four-nucleon coupling $\bar C_0$. 
For all these sources, we compare the deuteron TQM 
to its MQM, 
given by 
\cite{Vri11b}
\begin{equation}\label{MQMscale}
\mathcal M_d =\frac{ e g_A \bar g_0}{m_{\pi}} \frac{1}{2\pi F^2_{\pi}} 
\left[ (1+\kappa_0) + \frac{\bar g_1}{3 \bar g_0} (1+\kappa_1)\right] \, 
\frac{1+\xi}{(1+2\xi)^2} + 
e (1+\kappa_0) \frac{\mu-\gamma}{2\pi}\bar C_0.
\end{equation}
We consider the dimensionless ratio
$F_{\pi} \mathcal T_d/\mathcal M_d$, which, by power counting, 
is expected to be of order $h_1/F_{\pi}$. 

For the $\bar\theta$ term, qCEDM, and FQLR, $\bar C_0$ is subleading in
Eq. \eqref{MQMscale}, so that from Eq. \eqref{Tdtheta}
\begin{equation}\label{ratiog0}
F_{\pi} \left| \frac{ \mathcal T_d }{ \mathcal M_d} 
\right|_{\bar\theta,\, \textrm{qCEDM},\, \textrm{FQLR}} \simeq
0.4 \left|\frac{\bar g_0}{ \bar g_0 + 1.8 \bar g_1}\right| \frac{|h_1|}{F_{\pi}}. 
\end{equation}
For the $\bar\theta$ term, one can neglect $\bar g_1$ and 
the $\slashPT$ couplings
drop out of the ratio, which is approximately $|h_1|/F_{\pi}$, 
as expected by power counting.
For the qCEDM  
the ratio in Eq. \eqref{ratiog0} depends on 
$|\bar g_0/\bar g_1|$, which by NDA is expected to be order one.
For the FQLR, as discussed in Ref. \cite{deVries:2012ab}, $\bar g_0$ 
is somewhat suppressed with respect to $\bar g_1$, 
further suppressing the deuteron TQM with respect to the MQM.
In the case of $\chi$ISs, 
 $\bar C_0$ is expected
to be the leading term in
Eq. \eqref{MQMscale};
if we neglect the contribution of the nucleon EDM in Eq. \eqref{TdchiISs},
we get
\begin{equation}\label{ratioC0}
F_{\pi}\left| \frac{ \mathcal T_d}{ \mathcal M_d} \right|_{\textrm{$\chi$IS}}
\simeq 0.2 \frac{|h_1|}{F_{\pi}},
\end{equation}
which is also in good agreement with the NDA expectation.

For the remaining dimension-six source,
the qEDM, 
$d_d$ is also well approximated by the isoscalar nucleon EDM, 
\begin{equation}
d_d = 2 \bar d_0, 
\end{equation}
while
${\mathcal M}_d$ is suppressed by one power of $Q/\MNN$ with respect to the 
EDM \cite{Vri11b}.
Therefore, for the qEDM we compare the deuteron TQM with its EDM 
using the dimensionless ratio $m_N F_{\pi} \mathcal T_d/d_d$.
{}From Eq. \eqref{TdqEDM},
\begin{equation}\label{ratioEDM}
m_N F_{\pi}\left| \frac{ \mathcal T_d}{d_d} \right|_{\textrm{qEDM}}
\simeq 0.03 \left|\frac{\bar d_1}{\bar d_0}\right| \frac{|h_1|}{F_{\pi}} ,
\end{equation}
which is a bit smaller than naively expected.

Equations \eqref{ratiog0}, \eqref{ratioC0} and \eqref{ratioEDM}	make it 
explicit that 
the 
deuteron TQFF, in natural units, is suppressed roughly by a factor of 
$h_1/F_{\pi} \sim G_F \MQCD^2/4\pi \sim 10^{-6}$ with respect to the largest 
$\slashPT$ moment.
The lack of any significant numerical enhancement thus leads
to a very small TQFF.
The  
bounds on $\bar g_0$, $\bar d_1$, and $h_1$ 
inferred
in Sec. \ref{interactions} allow us to estimate the size of the TQM. 
For the QCD $\bar\theta$ term, the qCEDM, and the FQLR we find 
\begin{equation}
| \mathcal T_d |_{\bar\theta,\, \textrm{qCEDM},\, \textrm{FQLR}} \lesssim 
1.2 \cdot 10^{-19}\, e \, \textrm{fm}^3,
\end{equation}
while for the qEDM we find the even smaller value
\begin{equation}
|\mathcal T_d |_{\textrm{qEDM}} \lesssim 3.5 \cdot 10^{-21}\, e \, \textrm{fm}^3.
\end{equation}
For $\chi$ISs, one expects a similar value, but to be more precise a bound 
on $\bar C_0$ is needed.

These estimates have been obtained in chiral EFT with perturbative pions.
Iterating pions one can extend the regime of validity of the theory
beyond $\MNN$ at the cost of much more complicated renormalization
\cite{nogga}. Because the binding momentum of nucleons in the deuteron
is $\gamma\ll \MNN$, we do not expect drastic changes in the quantities
calculated here. In the case of our comparison $\slashPT$ moments,
this expectation has been checked \cite{liu}
and shown to be reasonable.

We conclude that the value of the deuteron TQM from parity violation in the SM 
and parity- and
time-reversal violation due to the SM $\bar{\theta}$ term or 
dimension-six operators originating
beyond the SM is, not surprisingly, tiny.
Evidence for a nonzero value for the deuteron TQM
that is larger than the ``background'' value 
$\sim 10^{-19}\, e \, \textrm{fm}^3$ would
likely be due to new $P\slashT$ interactions.

\section*{Acknowledgements}
U. van Kolck acknowledges the hospitality of the KVI Groningen on many 
occasions. This research was supported 
by the Dutch Stichting FOM under programs 104 and 114 (JdV, RGET) and 
in part by the DFG and the NSFC through funds provided to
the Sino-German CRC 110 ``Symmetries and the Emergence of Structure in QCD''  
(JdV), 
by the US DOE under contract DE-AC02-05CH11231 with the Director, 
Office of Science, Office of High Energy Physics (EM), 
and under grant DE-FG02-04ER41338 (UvK),
and by the Universit\'e Paris Sud under the program 
Attractivit\'e 2013 (UvK).

\section*{Appendices}

\appendix

\section{Dimension-six operators}
\label{app:dimsix}

The $\slashPT$ operators of dimension four and 
six,
after integrating out physics at the scale $M_{\slashT}$ and the heavy degrees
of freedom in the SM, are given by \cite{deVries:2012ab}
\begin{eqnarray}\label{QCDscale}
\mathcal L_{\slashPT} &=& m_{\star} \bar\theta \bar q i \gamma_5 q
-\frac{1}{2}\qb \left(d_0+d_3 \tau_3\right)i \simu \g_5 q \; \Fmu 
-\frac{1}{2}\qb \left(\tilde{d}_0+\tilde{d}_3 \tau_3\right)
i \simu\g_5\la^a q \; \Gmu\nonumber\\
&&
+ \frac{d_{W}}{6} f^{a b c} \ep^{\mu \nu \alpha \beta} 
G^a_{\alpha \beta} G_{\mu \rho}^{b} G^{c\, \rho}_{\nu} \nonumber\\
&& 
+ \frac{\textrm{Im}\,\Xi_1}{4} \ep^{3ij} \bar q \tau^i \gamma^{\mu}q \, 
\bar q \tau^j \gamma_{\mu} \gamma_5 q  
+ \frac{\textrm{Im}\,\Xi_8}{4} \ep^{3ij} \bar q \tau^i  \gamma^{\mu}\lambda^a q \, 
\bar q \tau^j  \gamma_{\mu} \lambda^a \gamma_5 q
\nonumber\\
&&
+ \frac{\textrm{Im}{\Sigma_1}}{4}  \left( \bar q q\, \bar q i \gamma_5 q 
- \bar q \boldtau q\, \cdot \bar q \boldtau i \gamma_5 q \right)
+  \frac{\textrm{Im}{\Sigma_8}}{4}  
\left( \bar q \lambda^a q\, \bar q i \gamma_5 \lambda^a q 
- \bar q \boldtau \lambda^a q\, \cdot \bar q \boldtau i \gamma_5 \lambda^a q 
\right).
\end{eqnarray}
The first operator is the QCD $\bar\theta$ term, where $m_{\star}$ denotes the
reduced light-quark mass $m_{\star} = m_u m_d/(m_u+m_d)$. We assume that
$\bar\theta \ll 1$. 
In the second and third operators, $d_0$ ($d_3$)
and $\tilde d_0$ ($\tilde d_3$) are the isoscalar (isovector) components of the 
quark EDM (qEDM) and quark chromo-EDM (qCEDM), respectively.
In the fourth term, $d_W$ represents
the gluon chromo-EDM (gCEDM).
The remainder 
consists of $\slashPT$ four-quark operators. 
The ones with coefficients $\Sigma_{1,8}$ are invariant under the SM gauge 
group, 
and can be generated directly at the electroweak scale.
The
isospin-breaking four-quark operators (FQLR) with coefficients
$\Xi_{1,8}$, on the other hand, are generated by
integrating out the weak gauge bosons and running to low energy.

The importance of the dimension-six $\slashPT$ operators depends on the high 
energy scale $M_{\slashT}$, on the detailed mechanism of $P$ and $T$ breaking 
in new physics, and
on the running to the QCD scale 
(for the latter, see Ref. \cite{Dekens:2013zca} and references therein).
We hide all the model dependence by introducing the dimensionless parameters
$\delta_{0,3}$, $\tilde \delta_{0,3}$, $w$, $\xi$, and $\sigma_{1,8}$
for (isoscalar and isovector) qEDM and qCEDM, gCEDM, and isospin-breaking
and invariant four-quark operators, respectively.  
We write \cite{deVries:2012ab}
\begin{eqnarray}
&&d_{0,3} =\Or\!\left(\frac{e\delta_{0,3} \bar m}{M^2_{\slashT}}\right),
\qquad
\tilde d_{0,3} =
\Or\!\left(4\pi \frac{\tilde\delta_{0,3}\bar m}{M^2_{\slashT}}\right),
\qquad 
d_W = \Or\!\left(4\pi\frac{w}{M^2_{\slashT}}\right),
\nonumber\\
&&\Xi_{1,8} = \Or\!\left(\frac{(4\pi)^2\xi}{M^2_{\slashT}}\right),
\qquad
\Sigma_{1,8} = \Or\!\left(\frac{(4\pi)^2\sigma_{1,8}}{M^2_{\slashT}}\right).
\label{scalingofdim6}
\end{eqnarray}
Naively,
one expects $\delta_{0,3}$, $\tilde{\delta}_{0,3}$, $w$, $\xi$, and 
$\sigma_{1,8}$ to be
$\Or(1)$, $\Or(g_s/4\pi)$, $\Or((g_s/4\pi)^3)$, $\mathcal O(1)$, and 
$\mathcal O(1)$, respectively,
but they could be significantly smaller or larger, 
depending on the new-physics model. 
As in the case of electroweak interactions, the relative strength of 
$\slashPT$ interactions
at low energies is expressed by the ratio of a low-energy scale and the 
characteristic scale
where $P$ and $T$ violation arise. The dimension-four $\bar\theta$ term 
is not suppressed
by any high energy scale, and its reduced coupling is  
$\bar \theta m^2_{\pi}/\MQCD^2$, where
the pion mass is a reminder of the intimate relationship 
between the $\bar\theta$ term and the quark masses.
In the case of dimension-six operators, the qEDM and qCEDM require 
an insertion of the
Higgs vacuum expectation value, which can be traded for the quark mass.
A dimensionless measure of their importance is then 
$(\tilde\delta_{0,3}, \delta_{0,3}) m^2_{\pi}/M^2_{\slashT}$.
For the remaining dimension-six $\slashPT$ operators the relevant low energy 
scale is $\MQCD$,
and the reduced couplings are $(w,\sigma_{1,8},\xi) \MQCD^2/M^2_{\slashT}$.

\section{Orders of magnitude}
\label{app:ordmag}

In terms of the dimensionless parameters defined in App. \ref{app:dimsix}, 
we can
estimate \cite{BiraEmanuele,deVries:2012ab} 
the size of the couplings $\bar g_0$ and $\bar d_1$ in 
Eq. \eqref{g0Lag} using NDA as
\begin{eqnarray}
\bar g_0 &=& \mathcal O\left(\bar\theta \frac{m^2_{\pi}}{\MQCD},\, 
(\tilde\delta_0,\varepsilon\tilde\delta_3)\frac{m^2_{\pi}\MQCD}{M^2_{\slashT}},\, 
\varepsilon \xi \frac{\MQCD^3}{M^2_{\slashT}},\, 
(w,\sigma_{1,8}) \frac{m^2_{\pi} \MQCD}{M^2_{\slashT}},\, 
\delta_{0,3}\frac{\alpha_{\mathrm{em}}}{4\pi}\frac{m^2_{\pi}\MQCD}{M^2_{\slashT}}
\right) ,
\label{g0scale}
\\
\frac{\bar d_1}{e}&=& \mathcal O\left(
\bar\theta \frac{m^2_{\pi}}{\MQCD^3},\, 
\tilde\delta_{0,3} \frac{m^2_{\pi}}{\MQCD M^2_{\slashT}},\, 
\xi \frac{\MQCD}{M^2_{\slashT}},\, 
(w,\sigma_{1,8}) \frac{\MQCD}{M^2_{\slashT}},\, 
(\varepsilon \delta_0\frac{m^2_{\pi}}{\MQCD^2}, \delta_3) 
\frac{m^2_{\pi}}{\MQCD M^2_{\slashT}}
\right),
\label{d1scale}
\end{eqnarray}
Here $\varepsilon$ is related to the light quark mass difference by 
$m_d-m_u\equiv \varepsilon(m_u+m_d)$.
In the case of the QCD $\bar\theta$ term, chiral symmetry relates $\bar g_0$  
to the strong contribution to
the neutron-proton mass difference  by \cite{CDVW79, BiraEmanuele}
$\bar g_0 = \delta m_N (1-\varepsilon^2)\bar\theta/2\varepsilon
\simeq 3 \bar\theta$ MeV,
using results from a lattice QCD calculation \cite{latticedeltamN}.
For the dimension-six operators, going beyond NDA requires additional input 
from lattice QCD
or other non-perturbative techniques. The contributions to $\bar g_0$ from the 
isovector component of the qCEDM and the FQLR, which break isospin, 
are suppressed 
by the quark-mass difference. 
These two sources, as well as 
the $\chi$ISs, 
generate at leading order
also the isovector pion-nucleon coupling $\bar g_1$.
However, as we discuss in the main text,
such a coupling does not contribute to the TQFF at LO. 
For all sources there is also an isoscalar EDM component $\bar d_0$,
but this term is not relevant here either. 

The estimate of the LEC $\bar C_0$ associated with the 
four-nucleon operator in 
Eq. \eqref{g0Lag} requires more care \cite{Vri11b, JordyThesis}. 
In the EFT where pion exchange is treated perturbatively, pions with momentum 
above $\MNN$ are integrated out, which induces contributions to multi-nucleon 
interactions. In particular, $\bar C_0$  can be generated by a high-energy 
pion exchange between two nucleons with one vertex originating in $g_A$ and 
the other in $\bar g_0$. For $\chi$ISs a larger contribution to $\bar C_0$ is 
generated if one uses, instead of $\bar g_0$, 
$(\bar \zeta_1/F_\pi) (D^2 \boldpi)\cdot \bar N \boldtau N$,
where $\bar \zeta_1 = \Or ((w, \sigma_{1,8})\MQCD/M^2_{\slashT})$ 
\cite{deVries:2012ab}. 
Application of the NDA rules at the scale $\MNN$ then gives 
$\bar C_0 = \Or (g_A \bar g_0\,4\pi/m_N M_{N\!N}^3) $ or 
$\bar C_0 = \Or (g_A \bar \zeta_1\,4\pi/m_N M_{N\!N}) $. 
However, this naive scaling is altered because the operator connects 
to an $S$ wave. As discussed in Ref. \cite{ksw}, this enhances the scaling 
by a factor $\MNN/Q$ due to non-perturbative renormalization by the 
leading-order $PT$ two-nucleon interaction. One finds
\begin{equation}
\frac{m_N \gamma}{4\pi}\bar C_{0} = 
\mathcal O\left( 
\bar\theta \frac{m^2_{\pi}}{\MNN^2 \MQCD},\, 
(\tilde\delta_0,\varepsilon\delta_3)
\frac{m^2_{\pi}\MQCD}{\MNN^2 M^2_{\slashT}},\,
\xi \frac{\MQCD^3}{\MNN^2 M^2_{\slashT}}, \,
(w,\sigma_{1,8}) \frac{\MQCD}{M^2_{\slashT}},\, 
\delta_{0,3}\frac{\alpha_{\mathrm{em}}}{4\pi}
\frac{m^2_{\pi}\MQCD}{\MNN^2 M^2_{\slashT}}
\right).
\end{equation}

Combining the scaling of these low-energy constants 
with the power-counting rules outlined in Sec. \ref{calculation}, 
we find for the size
of the diagrams in Figs. \ref{Fig1}, \ref{Fig2},
and \ref{Fig3}:
\begin{eqnarray}
\textrm{Fig. \ref{Fig1}}&=&
\mathcal O\left(\frac{eG_{F}}{4\pi \MNN}\right)
\times\
\nonumber\\
&&
\mathcal O\left(
\bar\theta,\,  
(\tilde\delta_0,\varepsilon \tilde\delta_3)  \frac{\MQCD^2}{M^2_{\slashT}}, \,
\varepsilon \xi \frac{\MQCD^4 }{ Q^2 M^2_{\slashT}} ,\, 
(w,\sigma_{1,8}) \frac{\MQCD^2}{M^2_{\slashT}},\, 
\delta_{0,3} \frac{\alpha_{\mathrm{em}}}{4\pi} \frac{\MQCD^2}{M^2_{\slashT}}
\right),
\label{Fig1scale}
\\
\textrm{Fig. \ref{Fig2}}&=& 
\mathcal O\left( \frac{eG_F Q}{4\pi \MNN^2} \right)
\times\
\nonumber\\
&&
\mathcal O\left( 
\bar\theta ,  \,  
(\tilde\delta_0,\varepsilon \tilde\delta_3) \frac{\MQCD^2}{M^2_{\slashT}},\,
\varepsilon \xi \frac{\MQCD^4 }{ Q^2 M^2_{\slashT}}, \,
(w,\sigma_{1,8}) \frac{\MNN^2\MQCD^2}{Q^2M^2_{\slashT}}, \, 
\delta_{0,3} \frac{\alpha_{\mathrm{em}}}{4\pi}\frac{\MQCD^2}{M^2_{\slashT}}
\right),
\label{Fig2scale}
\\
\textrm{Fig. \ref{Fig3}(a)}&=&\mathcal O\left(\frac{eG_{F}}{4\pi \MQCD}\right)
\times\
\nonumber\\
&&\mathcal O\left(
\bar\theta,\,  
(\tilde\delta_0,\varepsilon \tilde\delta_3)  \frac{\MQCD^2}{M^2_{\slashT}}, \,
\varepsilon \xi \frac{\MQCD^4 }{Q^2 M^2_{\slashT}} ,\, 
(w,\sigma_{1,8}) \frac{\MQCD^2}{M^2_{\slashT}},\, 
\delta_{0,3} \frac{\alpha_{\mathrm{em}}}{4\pi} \frac{\MQCD^2}{M^2_{\slashT}}
\right),
\label{Fig3ascale}
\\
\textrm{Fig. \ref{Fig3}(b)}&=& 
\mathcal O\left( \frac{eG_F Q}{4\pi \MNN \MQCD} \right)
\times\
\nonumber\\
&&
\mathcal O\left( 
\bar\theta ,  \,  
(\tilde\delta_0,\varepsilon \tilde\delta_3) \frac{\MQCD^2}{M^2_{\slashT}},\,
\varepsilon \xi \frac{\MQCD^4 }{Q^2M^2_{\slashT}}, \,
(w,\sigma_{1,8}) \frac{\MNN^2\MQCD^2}{Q^2M^2_{\slashT}}, \, 
\delta_{0,3} \frac{\alpha_{\mathrm{em}}}{4\pi}\frac{\MQCD^2}{M^2_{\slashT}}
\right),
\label{Fig3bscale} \\
\textrm{Fig. \ref{Fig3}(c)}&=& 
\mathcal O\left(\frac{eG_FQ}{4\pi \MQCD^2} \right)
\times\
\nonumber\\
&&
\mathcal O\left( 
\bar\theta, \,
\tilde\delta_{0,3} \frac{\MQCD^2}{M^2_{\slashT}},\,
\xi \frac{\MQCD^4}{Q^2 M^2_{\slashT}} ,\,
(w,\sigma_{1,8}) \frac{\MQCD^4}{Q^2 M^2_{\slashT}}\,, 
(\varepsilon \delta_0\frac{m^2_{\pi}}{\MQCD^2}, \delta_3)
\frac{\MQCD^2}{M^2_{\slashT}}  
\right).
\label{Fig3cscale}
\end{eqnarray}
Various statements made in the text about relative magnitudes
of the $\slashPT$ sources follow straghtforwardly from these relations.

The same NDA technique  
can be used to estimate the size of other $\slashPT$ contributions.
For example, the contributions of 
short-range
$\slashPT$ nucleon-nucleon-photon interactions
to the $P\slashT$ form factor of the deuteron are,
for all sources, subleading with respect to 
Figs. \ref{Fig1}, \ref{Fig2}, and \ref{Fig3}.
Likewise,
short-range $\slashP T$ nucleon-nucleon interactions contribute to the 
$\slashP T$ two-nucleon potential at higher
order than one-pion exchange diagrams involving $h_1$ \cite{Zhu:2004vw}. 
As a consequence, diagrams involving
these interactions are also suppressed compared to the diagrams discussed here. 
The same holds for diagrams with
$\slashP T$ nucleon-nucleon-photon ($N\!N\!N\!N\gamma$) interactions \cite{springer}.

In addition, one can understand why 
effective $P\slashT$ interactions contribute at higher orders.
One-pion exchange with a $P\slashT$ coupling does not contribute to the 
$P\slashT$ potential in
the two-nucleon sector \cite{simonius}. 
The latter is dominated by exchanges of heavier mesons, like
the $\rho$ and the $a_1$, which, at low energy, appear as two-nucleon 
contact interactions with at
least two derivatives (see, {\it e.g.}, Ref. \cite{Song:2011sw}). 
The size of these two-nucleon operators can be again estimated in
NDA by multiplying
the reduced couplings for the $\slashP T$ and $\slashPT$ interactions. 
For example, for the $\tb$ term a $P\slashT$
two-nucleon coupling 
would scale as 
$\tb G_F Q/(4\pi m_N \MQCD^2)$.
Inserting this coupling into a two-loop diagram with the photon interacting 
on a nucleon line via the nucleon magnetic
moment gives rise to a TQM of order 
$\mathcal T_d =\Or(e\bar\theta G_F Q^2/((4\pi)^2\MQCD^3))$. 
Such a
contribution is much smaller than the long-range component that results 
from the pion-nucleon
couplings $h_1$ and $\bar g_0$ in Fig. \ref{Fig1}, 
as can be seen from comparison with Eq. \eqref{Fig1scale}. 
Similar power-counting estimates indicate that contributions from
diagrams with short-range $P\slashT$ $N\!N\!N\!N\gamma$ vertices
are smaller than the contributions in Eqs. (\ref{Fig1scale})-(\ref{Fig3cscale})
by at least a factor $Q^2/M_{N\!N}^2$.

For dimension-six sources, we can neglect $P\slashT$ 
operators in the EFT as well. For non-electromagnetic sources such as 
the qCEDM, the FQLR operators, and the $\chi$ISs the argument is analogous 
to the one used for the $\tb$ term. 
For the qEDM one might think that $P\slashT$ pion-nucleon-photon interactions 
could be relevant. Indeed, a $P\slashT$ operator of the form
$D_\mu \boldpi\cdot \Nb S^\nu \boldtau N\,e\Fmu$
gives a nonzero contribution to the deuteron TQM via a two-loop diagram. 
However, by power counting we find this diagram
to be smaller by a factor $Q^2/\MQCD^2$ compared to the diagram in 
Fig. \ref{Fig3}, which provides the dominant contribution for the qEDM. 

\section{Form factor integrals}
\label{app:ffs}

In Sec. \ref{calculation} the results for the deuteron TQFF 
$F_{P \slashT}(\vec q\,^2)$ were
expressed in terms of
$L=1,2,3$ loop integrals $I^{(L)}(\xi, \vec x)$, 
where 
$\xi=\gamma/m_{\pi}$ and $\vec x = {\vec q}/(4\gamma)$. 
We list here the form and expansions of
these integrals to terms of order $x^2$, where $x = |\vec x|$.

The one-loop integral in Eq. \eqref{FF1}
is standard, appearing for example in the
deuteron magnetic quadrupole form factor \cite{Vri11b}. 
It has the simple closed form
\begin{equation}
I^{(1)}\left(\vec x\right)
=\frac{\arctan x}{x} \ .
\end{equation}

The integrals appearing in the two- and three-loop diagrams are
more complicated. We express them in terms 
of the dimensionless variables $\vec y_i$, obtained by rescaling the loop 
momenta, $\vec k_i = m_{\pi} \vec y_i$. 
They can be conveniently calculated in 
coordinate space \cite{Binger:1999rq,fms}.

The two-loop functions 
in Eqs. \eqref{I2ab}, \eqref{I2c}, and \eqref{nEDM} depend on 
three two-loop integrals,
\begin{eqnarray}
I_1^{(2)}(\xi,\vec x) &=& \frac{1}{x^2(2\pi)^4} \, 
\int d^3 y_1 \, d^3 y_2 \, \frac{\vec y_2 \cdot \vec x}{\vec y_2^{\, 2} + 1} \,
\frac{1}{\vec y_1^{\, 2} + \xi^2}  \,
\frac{1}{\left(\vec y_1 + \vec y_2 \right)^2 + \xi^2}  \,
\frac{1}{\left(\vec y_1 + 2\xi \vec x\right)^2 + \xi^2} ,
\\
I_2^{(2)}(\xi,\vec x) &=& \frac{1}{x^4(2\pi)^4} \, 
\int d^3 y_1  \, d^3 y_2  
\left( \vec y_2 \cdot \vec x \, \vec y_1 \cdot \vec x 
- \frac{x^{2}}{3} \vec y_1 \cdot \vec  y_2\right)
\frac{1}{\vec y_1^{\, 2} + \xi^2}  \,
\frac{1}{\left(\vec y_1 + \vec y_2 \right)^2 + \xi^2}
\nonumber \\ 
& & \qquad\qquad 
\frac{1}{\vec y_2^{\, 2} + 1}\,
\frac{1}{\left(\vec y_1 + 2\xi \vec x\right)^2 + \xi^2},
\\
I^{(2)}_3(\xi,\vec x) &=& \frac{1}{x^4(2\pi)^4} \, 
\int d^3 y_1  \, d^3 y_2  
\left( \vec y_2 \cdot \vec x \, \vec y_1 \cdot \vec x 
- \frac{x^{2}}{3} \vec y_1 \cdot \vec  y_2\right)
\frac{1}{\vec y_1^{\, 2} + \xi^2} \,
\frac{1}{(\vec y_1 + \vec y_2)^2 + \xi^2} 
\nonumber \\ 
&& \qquad\qquad 
\frac{1}{\left(\vec y_2 + 2 \xi \vec x \right)^{\, 2} + 1} \,
\frac{1}{\left(\vec y_2 - 2 \xi \vec x \right)^{\, 2} + 1}.
\end{eqnarray}
Expanding in $\vec x^{\,2}$ and retaining the first two terms in the expansion,
we find
\begin{eqnarray}
I_1^{(2)}(\xi,\vec x) &=& \frac{1+\xi}{12(1+2\xi)^2} 
- x^2 \frac{10 + 65 \xi + 144 \xi^2+72 \xi^3}{360 (1 + 2\xi)^4} 
+ \mathcal O\left( x^4 \right),
\\
I_2^{(2)}(\xi,\vec x) &=& - \frac{10 + 27 \xi + 18 \xi^2}{540 (1+ 2\xi)^3} 
+ x^2 \frac{70 + 595 \xi + 1918 \xi^2+ 2400 \xi^3+ 960 \xi^4}{12600 (1+2\xi)^5} 
+ \mathcal O\left( x^4 \right), 
\\
I_3^{(2)}(\xi,\vec x) &=& - \frac{\xi (4+21 \xi + 30 \xi^2)}{540 (1+2\xi)^3} 
+ x^2 \frac{ \xi^3
\left(68 + 590 \xi + 1820 \xi^2 + 2100 \xi^3 + 840 \xi^4\right)}{4725(1+ 2\xi)^5}
+ \mathcal O(x^4).
\nonumber\\
\end{eqnarray}
The two-loop functions $I^{(2)}_{a,b,c,d}$ are 
obtained from $I^{(2)}_{1,2,3}$ by multiplying them by prefactors that 
take into account spin and isospin factors, and the symmetry factor of each 
diagram:
\begin{eqnarray}
I^{(2)}_a(\xi,\vec x) &=&  12 I_2^{(2)}(\xi,\vec x) + 4 I_1^{(2)}(\xi,\vec x), \\
I^{(2)}_b(\xi,\vec x) &=&  24 I^{(2)}_3(\xi,\vec x), \\
I^{(2)}_c(\xi,\vec x) &=&  48 I_1^{(1)}(\xi,\vec x), \\
I^{(2)}_d(\xi,\vec x) &=& -  48 I_2^{(2)}(\xi,\vec x) - 8 I_1^{(2)}(\xi,\vec x). 
\end{eqnarray}

Similarly,
the three-loop functions that enter in Eq.~(\ref{I3ab}) are defined 
in terms of two three-loop tensor integrals,
\begin{eqnarray}
I_1^{(3)}(\xi,\, \vec x) &=&   \frac{1}{2\xi x^4 (2\pi)^6} 
\int d^3 y_1 \, d^3 y_2 \, d^3 y_3 
\left(\vec y_2\cdot\vec x \, \vec y_3 \cdot\vec x 
-\frac{x^{2}}{3}\vec y_2 \cdot \vec y_3   \right)  
 \frac{1}{\vec y_1^{\, 2} +\xi^2}\, 
\frac{1}{\vec y_3^{\, 2} + 1} 
\nonumber \\
& &\qquad\quad
\frac{1}{\left(\vec y_1 + \vec y_3\right)^2 + \xi^2} \, 
\frac{1}{\vec y_2^{\, 2} + 1}\, 
\frac{1}{\left(\vec y_1 + 2 \xi \vec x\right)^2 + \xi^2} \,
\frac{1}{\left(\vec y_1 + \vec y_2 + 2 \xi \vec x \right)^2 + \xi^2},
\\
I_2^{(3)}(\xi,\, \vec x) &=&  \frac{1}{2\xi x^4 (2\pi)^6} 
\int d^3y_1 \, d^3y_2 \, d^3 y_3 
\left( \vec y_2 \cdot \vec x \, \vec y_3 \cdot \vec x 
-\frac{x^{2}}{3} \vec y_2 \cdot \vec y_3   \right)  
\frac{1}{\vec y_1^{\, 2} +\xi^2}\,
\frac{1}{\vec y_3^{\, 2} + 1}
 \nonumber \\
& &\qquad\quad  
\frac{1}{\left(\vec y_1 + \vec y_3\right)^2 + \xi^2}\,
\frac{1}{\left( \vec y_2 + 2 \xi \vec x\right)^{\, 2}+ 1} \,
\frac{1}{\left( \vec y_2 - 2 \xi \vec x\right)^{\, 2}+ 1} \,
\frac{1}{\left(\vec y_1 + \vec y_2 \right)^2 + \xi^2} ,
\end{eqnarray} 
which are finite 
in three and four dimensions.
Again retaining terms up to order $\mathcal O(x^2)$, 
\begin{eqnarray}
I_1^{(3)}(\xi,\, \vec x) & =&  
\frac{ 15  + 75 \xi  + 110 \xi^2 + 30  \xi^3 - 12  \xi^4 + 8  \xi^5 - 48 \xi^6}
{2160 \xi^5  (1 + 2 \xi)^3} 
\log \left(\frac{2 (1+ \xi)}{1+ 2 \xi} \right)   
\nonumber \\ 
& & 
+ \frac{15  + 65 \xi + 75 \xi^2 + 7  \xi^3 - 30  \xi^4 - 12 \xi^5 - 24 \xi^6}
{4320 \xi^4 (1 + \xi) (1 + 2 \xi)^3}  
-\frac{1}{288 \xi^6} 
\left( \textrm{Li}_2\left( -\frac{1}{1+2\xi} \right) + \frac{\pi^2}{12}\right)
\nonumber \\
& & - x^2
\bigg[
\frac{1}{37800 \xi^5 (1+2\xi)^5}
\left(105  + 945 \xi + 3290 \xi^2 + 5390  \xi^3 + 3836 \xi^4 + 560  \xi^5 
\right. 
\nonumber \\  
& &  \left.
- 160 \xi^6 + 80 \xi^7 
- 4320 \xi^8  - 9024 \xi^9 - 5760 \xi^{10}\right) 
\log \left(\frac{2 (1+ \xi)}{1+ 2 \xi} \right)
\nonumber\\
&&
+ \frac{1}{75600 \,\xi^4 (1 + \xi)^3 (1 + 2 \xi)^5}
\left(105  + 1085 \xi + 4260 \xi^2 + 10374  \xi^3 + 12 789 \xi^4 
\right.
\nonumber\\
&&  \left.
+ 6765 \xi^5-5202 \xi^6 -10052 \xi^7 +5400 \xi^8 + 24384 \xi^9 + 20544 \xi^{10} 
+5760 \xi^{11}
\right)   
\nonumber\\
&& 
- \frac{1}{720 \xi^6} 
\left( \textrm{Li}_2\left( -\frac{1}{1+2\xi} \right)+\frac{\pi^2}{12}\right) 
\bigg] + \mathcal O( x^{\, 4})
\label{threel}
\end{eqnarray}
and
\begin{eqnarray}
 I^{(3)}_2(\xi,\vec x) &=& \frac{\xi}{540 (1+2\xi)^3} \left[ 
\frac{7 + 44 \xi + 58 \xi^2 + 20 \xi^3}{4 (1+\xi)^2} 
- (1+ 6 \xi) \log \left( \frac{2(1+\xi)}{1+2\xi}\right) \right]
\nonumber \\
& & + x^2 \frac{\xi^3}{90 (1+2\xi)^5} 
\left[\frac{5 + 50 \xi + 116 \xi^2 + 72 \xi^3}{3}
\log \left( \frac{2(1+\xi)}{1+2\xi} \right) 
\right. 
\nonumber\\ 
& & 
\left. + \frac{205 + 2494 \xi + 12290 \xi^2 + 29472 \xi^3 + 37540 \xi^4 
+ 25672 \xi^5 + 8720 \xi^6 + 1120 \xi^7}
{280 (1+\xi)^4}
\right]
\nonumber\\
&& + \mathcal O(x^4),
\end{eqnarray}
where $\textrm{Li}_2(x)$ is the dilogarithm function.
The function $I_2^{(3)}$ vanishes in the limit of vanishing binding momentum, 
$\xi \rightarrow 0$.
$I_1^{(3)}(\xi, 0)$ consists of the sum of three terms, each of them divergent 
for $\xi \rightarrow 0$. Because of cancellations between these terms, 
$I_1^{(3)}(\xi, 0)$ has a finite limit for vanishing $\xi$, 
$I_1^{(3)}(0,0) = -1/162$. However, this limit is not a good approximation to 
the value of the three-loop function at the physical value of $\xi$, 
$\xi \simeq 0.3$. One needs to keep at least terms up to $\xi^{15}$ to obtain 
an approximation of $I^{(3)}_1(0.3,0)$ at the $10 \%$ accuracy. 

As before, lumping spin, isospin, and symmetry factors in the definitions 
of $I^{(3)}_{a,b}$, we have
\begin{equation}
I^{(3)}_a (\xi, \vec x) = 36\,  I^{(3)}_1(\xi,\vec x), 
\qquad 
I^{(3)}_b (\xi, \vec x) = 240 \, I^{(3)}_2(\xi,\vec x). 
\end{equation}



\begin{thebibliography}{99}

\bibitem{Khr97}
I.B. Khriplovich and S.K. Lamoreaux,
\textit{CP Violation Without Strangeness: Electric Dipole Moments
of Particles, Atoms, and Molecules\/} (Springer Verlag, Berlin, 1997).

\bibitem{dianapole}
Ya.B. Zel'dovich,
Sov. Phys. JETP {\bf 6} (1958) 1184;
{\bf 12} (1961) 777.

\bibitem{quadruanapole}
V. Glaser and B. Ja\v ksi\'c,
Nuovo Cim. {\bf 5} (1957) 76;
I.Yu. Kobzarev, L.B. Okun', and M.V. Terent'ev,
JETP Lett. {\bf 2} (1965) 289;
A.D. Dolgov,
JETP Lett. {\bf 2} (1965) 308.

\bibitem{Gra10}
C.G. Gray, G. Karl, and V.A. Novikov,
Am. J. Phys. {\bf 78} (2010) 936.

\bibitem{boudjema}
F. Boudjema and C. Hamzaoui,
Phys. Rev. D {\bf 43} (1991) 3748.

\bibitem{atomic}
S.G. Porsev,
Phys. Rev. A {\bf 49} (1994) 5105.

\bibitem{edscatt}
V.M. Dubovik and A.A. Cheshkov,
Sov. Phys. JETP {\bf 24} (1967) 111;
V.M. Dubovik, E.P. Likhtman, and A.A. Cheshkov,
Sov. Phys. JETP {\bf 25} (1967) 464;
D. Schildknecht,
Z. Phys. {\bf 201} (1967) 99;
R. Prepost, R.M. Simonds, and B.H. Wiik,
Phys. Rev. Lett. {\bf 21} (1968) 1271;
K.Y. Lin,
Nucl. Phys. B {\bf 18} (1970) 162;
H. Arenh\"ovel and S.K. Singh,
Eur. Phys. J. A {\bf 10} (2001) 183.

\bibitem{COSY}
P.D. Eversheim, B. Lorentz, and Yu. Valdau (spokespersons),
Test of Time-Reversal Invariance in Proton-Deuteron Scattering at COSY,
http://apps.fz-juelich.de/pax/paxwiki/index.php/Test\_of\_Time-Reversal\_Invariance\_at\_COSY\_(TRIC);
P.D. Eversheim {\it et al.}, Hyperfine Interact. {\bf 193} (2009) 335;
Yu. Valdau, POS {\bf STORI11} 013;
M. Beyer, Nucl. Phys. A {\bf 560} (1993) 895.

\bibitem{Jarlskog}
C. Jarlskog,
Phys. Rev. Lett. {\bf 55} (1985) 1039.

\bibitem{'tHooft}
G. 't Hooft,
Phys. Rev. Lett. {\bf 37} (1976) 8;
Phys. Rev.  D {\bf 14} (1976) 3432,
{\it ibid.} {\bf 18} (1978) 2199(E).

\bibitem{dnbound}
C.A. Baker {\it et al.}, 
Phys. Rev. Lett. {\bf 97} (2006) 131801.

\bibitem{PslashT}
J. Engel, P.H. Frampton, and R.P. Springer,
Phys. Rev. D {\bf 53} (1996) 5112;
M.J. Ramsey-Musolf,
Phys. Rev. Lett.  {\bf 83} (1999) 3997,
{\it ibid.} {\bf 84} (2000) 5681(E);
A. Kurylov, G.C. McLaughlin, and M.J. Ramsey-Musolf,
Phys. Rev. D {\bf 63} (2001) 076007.

\bibitem{paulo}
P.F. Bedaque and U. van Kolck,
Ann. Rev. Nucl. Part. Sci. {\bf 52} (2002) 339.

\bibitem{veronique}
V. Bernard, N. Kaiser, and U.-G. Mei\ss ner,
Int. J. Mod. Phys. E {\bf 4} (1995) 653.

\bibitem{maekawa}
C.M. Maekawa and U. van Kolck, 
Phys. Lett. B {\bf 478} (2000) 73;
C.M. Maekawa, J.S. Veiga, and U. van Kolck, 
Phys. Lett. B {\bf 488} (2000) 167.

\bibitem{BiraHockings} 
W.H. Hockings and U. van Kolck, 
Phys. Lett. B {\bf 605} (2005) 273;
K. Ottnad, B. Kubis, U.-G. Mei{\ss}ner, and F.-K. Guo,
Phys. Lett. B {\bf 687} (2010) 42;
J. de Vries, E. Mereghetti, R.G.E. Timmermans, and U. van Kolck,
Phys. Lett. B {\bf 695} (2011) 268; 
E. Mereghetti, J. de Vries, W.H. Hockings, C.M. Maekawa, and U. van Kolck,
Phys. Lett. B {\bf 696} (2011) 97.

\bibitem{ksw}
D.B. Kaplan, M.J. Savage, and M.B. Wise,
Nucl. Phys. B {\bf 534} (1998) 329.

\bibitem{fms}
S. Fleming, T. Mehen, and I.W. Stewart,
Nucl. Phys. A {\bf 677} (2000) 313.

\bibitem{nogga}
S.R. Beane, P.F. Bedaque, M.J. Savage, and U. van Kolck,
Nucl. Phys. A {\bf 700} (2002) 377;
A. Nogga, R.G.E. Timmermans, and U. van Kolck,
Phys. Rev. C {\bf 72} (2005) 054006;
M. Birse, 
Phys. Rev. C {\bf 74} (2006) 014003;
M. Pav\'on Valderrama, 
Phys. Rev. C {\bf 83} (2011) 024003;
{\bf 84} (2011) 064002;
B. Long and C.J. Yang, 
Phys. Rev. C {\bf 84} (2011) 057001;
{\bf 85} (2012) 034002;
{\bf 86} (2012) 024001.

\bibitem{deutEMFF}
D.B. Kaplan, M.J. Savage, and M.B. Wise,
Phys. Rev. C {\bf 59} (1999) 617.

\bibitem{Vri11b}
J. de Vries, E. Mereghetti, R.G.E. Timmermans, and U. van Kolck,
Phys. Rev. Lett. {\bf 107} (2011) 091804.

\bibitem{springer}
M.J. Savage and R.P. Springer,
Nucl. Phys. A {\bf 686} (2001) 41.

\bibitem{Weinberg} 
S. Weinberg, 
\textit{The Quantum Theory of Fields\/}, Vol. 2
(Cambridge University Press, Cambridge, 1996).

\bibitem{NDA}
A.V. Manohar and H. Georgi, 
Nucl. Phys. B {\bf 234} (1984)  189;
H. Georgi and L. Randall,
Nucl. Phys. B {\bf 276} (1986) 241.

\bibitem{Kaplan:1992vj} 
D.B. Kaplan and M.J. Savage,
Nucl. Phys. A {\bf 556} (1993) 653;
{\it ibid.} {\bf 570} (1994) 833(E); 
{\bf 580} (1994) 679(E).

\bibitem{Zhu:2004vw}
S.L. Zhu, C.M. Maekawa, B.R. Holstein, M.J. Ramsey-Musolf, and U. van Kolck,
Nucl. Phys.  A {\bf 748} (2005) 435.

\bibitem{npdgammath}
D.B. Kaplan, M.J. Savage, and R.P. Springer,
Phys. Lett. B {\bf 449} (1999) 1.

\bibitem{npdgammaexp}
M.T. Gericke {\it et al.},
Phys. Rev.  C {\bf 83} (2011) 015505.

\bibitem{hpilatt}
J. Wasem,
Phys. Rev.  C {\bf 85} (2012) 022501.

\bibitem{Buchmuller:1985jz}
W. Buchm\"uller and D. Wyler,
Nucl. Phys. B {\bf 268} (1986) 621;
B. Grzadkowski, M. Iskrzynski, M. Misiak, and J. Rosiek,
JHEP {\bf 1010} (2010) 085.

\bibitem{deVries:2012ab}
J. de Vries, E.~Mereghetti, R.G.E. Timmermans, and U. van Kolck,
Ann. Phys. (to appear), {\tt arXiv:1212.0990 [hep-ph]}.

\bibitem{BiraEmanuele} 
E. Mereghetti, W.H. Hockings, and U. van Kolck,
Ann. Phys. {\bf 325} (2010) 2363.

\bibitem{CDVW79}
R.J. Crewther, P. Di Vecchia, G. Veneziano, and E. Witten,
Phys. Lett. B {\bf 88} (1979) 123; 
{\bf 91} (1980) 487(E).

\bibitem{simonius}
M. Simonius,
Phys. Lett. B {\bf 58} (1975) 147. 

\bibitem{liu}
J. de Vries, R. Higa, C.-P. Liu, E. Mereghetti, I. Stetcu, R.G.E. Timmermans,
and U. van Kolck,
Phys. Rev. C {\bf 84} (2011) 065501;
C.-P. Liu, J. de Vries, E. Mereghetti, R.G.E. Timmermans, and U. van Kolck,
Phys. Lett. B {\bf 713} (2012) 447.

\bibitem{Dekens:2013zca}
W. Dekens and J. de Vries,
JHEP (to appear),
{\tt arXiv:1303.3156 [hep-ph]}.

\bibitem{latticedeltamN}
S.R. Beane, K. Orginos, and M.J. Savage,
Nucl. Phys. B {\bf 768} (2007) 38.

\bibitem{JordyThesis}
J. de Vries, 
Ph.D. dissertation, University of Groningen (2012).

\bibitem{Song:2011sw}
Y.H.~Song, R.~Lazauskas, and V.~Gudkov,
Phys. Rev. C {\bf 83} (2011) 065503.

\bibitem{Binger:1999rq} 
M. Binger,
{\tt nucl-th/9901012}.

\end{thebibliography}
\end{document}